\documentclass[notitlepage,aps,prd,10pt,preprintnumbers,tightenlines,superscriptaddress,longbibliography,nofootinbib]{revtex4-1}
\usepackage[utf8]{inputenc}
\usepackage{amssymb,latexsym}
\usepackage{amsmath,amsbsy,bbm}
\usepackage{epsfig,bm}
\usepackage{graphicx,comment,color}
\usepackage{slashed}
\usepackage{array}

\begin{document}

%%%%%G%R%E%E%K%%%%%%%%%%%%
\def\a{{\alpha}}
\def\b{{\beta}}
\def\d{{\delta}}
\def\D{{\Delta}}
\def\X{{\Xi}}
\def\e{{\varepsilon}}
\def\g{{\gamma}}
\def\G{{\Gamma}}
\def\l{{\lambda}}
\def\L{{\Lambda}}
\def\m{{\mu}}
\def\n{{\nu}}
\def\o{{\omega}}
\def\O{{\Omega}}
\def\S{{\Sigma}}

%%%%%%%%%%%%%%%%%%%%%%%%%%%%%%%%%%%%%
\newcommand\beq{\begin{eqnarray}}
\newcommand\eeq{\end{eqnarray}}
\newcommand{\fr}[2]{{\frac{#1}{#2}\,}}
\newcommand{\nn}{\nonumber}

%%%%%%%%%%%%%%%%%%%%%%%%%%%%%%%%%%%%%
\newcommand{\Tr}{{\rm Tr} }
\newcommand\numberthis{\addtocounter{equation}{1}\tag{\theequation}}
\newcolumntype{L}{>{$}l<{$}}
\newcolumntype{R}{>{$}r<{$}}
\newcolumntype{C}{>{$}c<{$}}

%%%%%%%%%%%%%%%%%%%%%%%%%%%%%%%%%%%%%
\def\ol#1{{\overline{#1}}}
\def\c#1{{\mathcal #1}}
\def\eqref#1{{(\ref{#1})}}

%%%%%%%%%%%%%%%%%%%%%%%%%%%%%%%%%%%%%%%% 
\title{Octet Baryons in Large Magnetic Fields}
%%%%%%%%%%%%%%%%%%%%%%%%%%%%%%%%%%%%%%%% 

%%%%%%%%%%%%%%%%%%%%%%%%%%%%%%%%%%%%%%%% 
\author{Amol %A.~%
Deshmukh}
\email[]{$\texttt{adeshmukh@gradcenter.cuny.edu}$}
\affiliation{
Department of Physics,
        The City College of New York,
        New York, NY 10031, USA}
\affiliation{
Graduate School and University Center,
        The City University of New York,
        New York, NY 10016, USA}
%%%%%%%%%%%%%%%%%%%%%%%%%%%%%%%%%%%%%%%% 
\author{Brian~C.~Tiburzi}
\email[]{$\texttt{btiburzi@ccny.cuny.edu}$}
\affiliation{
Department of Physics,
        The City College of New York,
        New York, NY 10031, USA}
\affiliation{
Graduate School and University Center,
        The City University of New York,
        New York, NY 10016, USA}

%%%%%%%%%%%%%%%%%%%%%%%%%%%%%%%%%%%%%%%% 
\date{\today}

%\pacs{12.39.Hg, 13.40.Gp, 13.60.Fz, 14.20.Dh}

%%%%%%%%%%%%%%%%%%%%%%%%%%%%%%%%%%%%%%%% 
\begin{abstract}

Magnetic properties of octet baryons are investigated within the framework of chiral perturbation theory. 
Utilizing a power counting for large magnetic fields, 
the Landau levels of charged mesons are treated exactly giving rise to baryon energies that depend non-analytically on the strength of the magnetic field. 
In the small-field limit, 
baryon magnetic moments and polarizabilities emerge from the calculated energies. 
We argue that the magnetic polarizabilities of hyperons provide a testing ground for potentially large contributions from decuplet pole diagrams. 
In external magnetic fields, 
such contributions manifest themselves through decuplet-octet mixing, 
for which possible results are compared in a few scenarios. 
These scenarios can be tested with lattice QCD calculations of the octet baryon energies in magnetic fields. 

\end{abstract}
%%%%%%%%%%%%%%%%%%%%%%%%%%%%%%%%%%%%%%%% 

\maketitle

%%%%%%%%%%%%%%%%%%%%%%%%%%% 
\section{Introduction}                                                 %
%%%%%%%%%%%%%%%%%%%%%%%%%%%

Studying the response of systems to external conditions is a central theme
that appears in many branches of physics. 
In quantum field theory, 
the external field problem was pioneered long ago in the context of QED by Schwinger%
~\cite{Schwinger:1951nm}, 
yet, 
in a deeply insightful and modern way. 
For QCD, 
the color fields are confined within hadrons; 
but,
the quarks nonetheless carry charges that couple to other currents in the Standard Model.  
The QCD external field problem allows one to probe the rich behavior of strongly interacting systems under external conditions, 
including the modification of vacuum and hadron structure due to external electromagnetic fields. 
These dynamics, 
moreover, 
are likely relevant to describe the physics in the interiors of magnetars%
~\cite{Duncan:1992hi,Broderick:2001qw,Harding:2006qn}
and in non-central heavy-ion collisions%
~\cite{Skokov:2009qp,Kharzeev:2013jha,McLerran:2013hla}, 
for which large magnetic fields upwards of
$\sim 10^{19} \, \texttt{Gauss}$
are conceivable. 
A comprehensive overview of quantum field theories in external magnetic fields appears in
Ref.~\cite{Miransky:2015ava}.

While relevant in certain physical environments, 
the external field problem also provides a useful computational tool. 
For non-perturbative QCD calculations using lattice gauge theory, 
the external field technique has proven valuable. 
Uniform magnetic fields, 
for example, 
were employed in the very first lattice QCD computations of the nucleon magnetic moments%
~\cite{Bernard:1982yu,Martinelli:1982cb}. 
Since then,  
calculations continue to exploit features of the external field technique, 
such as: 
in computing electromagnetic polarizabilities%
~\cite{Fiebig:1988en,Christensen:2004ca,Lee:2005dq,Detmold:2009dx,Detmold:2010ts,Primer:2013pva,Lujan:2014kia,Freeman:2014kka,Luschevskaya:2014lga,Appelquist:2015zfa,Luschevskaya:2015cko}, 
which would otherwise require the determination of computationally expensive four-point correlation functions; 
and, 
in computing the magnetic properties of light nuclei%
~\cite{Beane:2014ora,Chang:2015qxa,Beane:2015yha}, 
for which even three-point correlation functions are not currently practicable for calculations. 
Additional studies explore the behavior of QCD in large magnetic fields, 
for example, 
the modification of nucleon-nucleon interactions%
~\cite{Detmold:2015daa}, 
and effects on the phase diagram of QCD%
~\cite{Bali:2011qj,Bali:2012zg,Bruckmann:2013oba,Bornyakov:2013eya,Bali:2014kia,Endrodi:2015oba}.

In this work, 
we explore the behavior of octet baryon energies in large magnetic fields. 
This investigation is carried out within the framework of chiral perturbation theory, 
which can be used to study, 
in a model-independent fashion, 
the modification of vacuum and hadron structure in large electromagnetic fields, 
see 
Refs.~\cite{Shushpanov:1997sf,Agasian:1999sx,Agasian:2001ym,Cohen:2007bt,Werbos:2007ym,Tiburzi:2008ma,Tiburzi:2014zva}. 
One motivation for this study is the prohibitive size of magnetic fields required in lattice QCD computations. 
Uniform magnetic fields on a torus are subject to 
't Hooft's quantization condition%
~\cite{tHooft:1979rtg}, 
which restricts such field strengths to satisfy 
$e B = 6 \pi \, n / L^2$, 
where 
$L$
is the spatial extent of the lattice, 
$n$ 
is an integer, 
and the factor of three arises from the fractional nature of the quark charges. 
Assuming 
$m_\pi L \sim 4$, 
the allowed magnetic fields satisfy 
$e B / m_\pi^2 \sim 1.2 \, n$. 
External field computations of hadron properties require several values of the magnetic field, 
moreover, 
leading us to consider 
$e B / m_\pi^2 \sim 3$--$4$
for the extraction of polarizabilities,
which enter as a quadratic response to the magnetic field.   
In this regime, 
chiral corrections from charged-pion loops are altered by Landau levels. 
The same is true for charged-kaon loops, 
although their alteration is comparatively less important. 
Both effects are addressed in the present work;%
\footnote{
In the power counting utilized below, 
the decuplet-octet mass splitting 
$\D$
is treated as the same order as the meson mass
$m_\phi$. 
Thus decuplet loop contributions, 
which introduce dependence on 
$e B / \D^2$,
are also treated non-perturbatively. 
} 
and, 
while they cannot be treated perturbatively in powers of the magnetic field, 
these effects lead to modifications of baryon energies that are nevertheless of a reasonably small size.

An additional feature appears in the magnetic-field dependence of octet baryon energies, 
namely that of mixing with decuplet baryons. 
Indeed, 
this complicating feature can be anticipated from studying the magnetic polarizabilities of the nucleon. 
The determination of nucleon electromagnetic polarizabilities using chiral perturbation theory has been the subject of continued effort%
~\cite{Bernard:1991rq,Butler:1992ci,Bernard:1993bg,Bernard:1993ry,Hemmert:1996rw,Hemmert:1997tj,McGovern:2001dd,Pascalutsa:2002pi,Beane:2004ra,Lensky:2009uv,McGovern:2012ew,Griesshammer:2015ahu}; 
and, 
there are a number of reviews focusing on different aspects of the subject%
~\cite{Drechsel:2002ar,Schumacher:2005an,Griesshammer:2012we,Holstein:2013kia,Hagelstein:2015egb}. 
The computation of nucleon electric polarizabilities has remained relatively uncontroversial, 
and the leading one-loop computation is already in good agreement with experimental determinations. 
The magnetic polarizabilities, 
by contrast,  
have proved challenging due to large paramagnetic contributions from the delta-pole diagram, 
and correspondingly large diamagnetic contributions from higher-order, 
short-distance operators. 
A central observation of the present work is that these contributions can be disentangled in external magnetic fields: 
the latter simply lead to energy shifts, 
while the former require the summation of 
$e B / (\D M_N)$
contributions, 
where 
$\D$
represents the delta-nucleon mass splitting, 
and
$M_N$
the nucleon mass. 
Such summation is achieved by diagonalizing the magnetically coupled 
delta-nucleon system. 
The role of loop contributions and decuplet mixing, 
moreover, 
is addressed within the entire baryon octet, 
for which 
$U$-spin and large-$N_c$ considerations allow us to compare results for magnetic polarizabilities 
and the behavior of energies with respect to the magnetic field. 
While the polarizabilities of hyperons have received comparatively less attention, 
see~\cite{Bernard:1992xi,Aleksejevs:2010zw,VijayaKumar:2011uw,Blin:2015era}, 
the lack of experimental constraints can be ameliorated with future lattice QCD computations. 
The results of such computations will enable paramagnetic and diamagnetic contributions to be disentangled, 
along with exposing the role symmetries play in the magnetic rigidity of baryons.

The organization of our presentation is as follows. 
First, 
in Sec.~\ref{s:chpt}, 
we review the necessary ingredients of meson and baryon chiral perturbation theory in large magnetic fields using a position-space formulation. 
Additionally we explain the partial resummations employed to mitigate the effects of 
$SU(3)_V$
breaking. 
Next, 
in Sec.~\ref{s:energy}, 
we determine expressions for the octet baryon energies as a function of the magnetic field to third order in the combined chiral and heavy baryon expansion. 
These results account for tree-level and loop contributions;
the former features a problematically large contribution from the decuplet pole diagram. 
The expressions for baryon energies are then utilized in Sec.~\ref{s:scenarios}, 
where three scenarios are investigated. 
We explore the likelihood that a large baryon transition moment leads to sizable mixing between decuplet and octet baryons in magnetic fields. 
Consistent kinematics are employed to reduce the size of magnetic polarizabilities,
as well as a scenario in which higher-order counterterms are promoted.  
These scenarios can be tested with future lattice QCD computations of the octet baryons in magnetic fields. 
In Appendix~\ref{s:magpol}, 
we provide the corresponding results for magnetic moments and electric polarizabilities computed in our approach. 
Technical details concerning the coupled three-state system of 
$I_3 = 0$
baryons are contained in 
Appendix~\ref{s:B}. 
Finally in 
Sec.~\ref{s:summy}, 
we conclude with a summary of our findings.

%%%%%%%%%%%%%%%%%%%%%%%%%%%%%%%%%%%%%%%
\section{Chiral Perturbation Theory in Large Magnetic Fields}                          %
%%%%%%%%%%%%%%%%%%%%%%%%%%%%%%%%%%%%%%%%
\label{s:chpt}

Our calculations of the octet baryon energies in large magnetic fields are performed using 
three-flavor chiral perturbation theory. 
Inclusion of the large magnetic field is achieved through a modified power-counting scheme. 
Here, 
we describe this scheme, 
as well as the necessary ingredients of meson and baryon chiral perturbation theory. 
The latter is implemented utilizing the heavy baryon framework.
Additionally, 
we employ a partial resummation of 
$SU(3)_V$
breaking effects.

%%%%%%%%%%%%%%%%%%%%%%%%%%%
\subsection{Meson Sector}

To compute the energies of the octet baryons, 
we consider the three-flavor chiral limit, 
$m_u=m_d=m_s=0$, 
about which an effective field theory description in terms of 
chiral perturbation theory
($\chi$PT) 
is possible.  
In this limit, 
the 
$SU(3)_L\times SU(3)_R$ 
chiral symmetry of QCD is spontaneously broken to 
$SU(3)_V$ 
by the formation of the quark condensate. 
The emergent Goldstone bosons, 
which are identified as the octet of pseudoscalar mesons 
($\pi$, $K$, $\eta$), 
are parameterized as elements of the coset space 
$SU(3)_L\times SU(3)_R/SU(3)_V$
in the form 
\beq
\Sigma=\exp\left(\fr{2i\phi}{f_\phi}\right),
\eeq
where
\beq
\phi^i_j=
  \begin{pmatrix}
    \frac{1}{\sqrt{2}}\pi^0+\frac{1}{\sqrt{6}}\eta & \pi^+ & K^+  \\
    \pi^- & -\frac{1}{\sqrt{2}}\pi^0+\frac{1}{\sqrt{6}}\eta & K^0  \\
    K^- & \ol K {}^0 & -\frac{2}{\sqrt{6}}\eta
  \end{pmatrix}^i_j.
\eeq
Strictly speaking, 
$f_\phi$ 
is the three-flavor chiral-limit value of the pseudoscalar meson decay constant. 
Due to rather large 
$SU(3)_V$ 
breaking effects, 
we employ differing decay constants within each meson isospin multiplet. 
This corresponds to a partial resummation of higher-order terms in the chiral expansion. 
With our normalization, 
the values for charged-meson decay constants obtained from experiment are 
$f_\pi=130~\verb|MeV|$ and $f_K=156~\verb|MeV|$ \cite{Olive:2016xmw}. 
Note that 
$f_\eta$ 
is not required in this work.

The low-energy dynamics of the pseudoscalar meson octet can be described in an effective field theory framework, 
which is three-flavor 
$\chi$PT%
~\cite{Gasser:1984gg}. 
The chiral Lagrangian density is constructed based on the pattern of explicit and spontaneous symmetry breaking. 
The sources of explicit symmetry breaking are the masses and the electric charges of the quarks, 
which are encoded in the matrices 
$m_q=\text{diag} \left(\ol m, \ol m,m_s \right)$ 
and  
$\c Q=\text{diag}\left(\frac{2}{3},-\frac{1}{3},-\frac{1}{3}\right)$, 
respectively. 
We work in the strong isospin limit, 
and accordingly use the isospin-averaged quark mass
$\ol m$,
which is given by
$\ol m = \frac{1}{2} (m_u + m_d)$. 
To organize the infinite number of possible terms in the chiral Lagrangian density, 
we assume the power counting
\beq
\label{eq:1}
\frac{k^2}{\Lambda_\chi^2}
\sim 
\frac{m_\phi^2}{\Lambda_\chi^2}
\sim
\frac{(eA_\mu)^2}{\Lambda_\chi^2}
\sim
\frac{eF_{\mu\nu}}{\Lambda_\chi^2}
\sim
\epsilon^2
,\eeq
where 
$k$ 
is the meson momentum, 
$m_\phi$ 
is the mass of the meson,  
$A_\mu$ 
is the electromagnetic gauge potential,  
$F_{\mu\nu}$ 
is the corresponding field-strength tensor, 
and 
$\epsilon$
is assumed to be small,
$\epsilon \ll 1$. 
Notice that in the computation of gauge-invariant quantities, 
$A_\mu$
cannot appear. 
The cutoff scale of the effective theory, 
$\L_\chi$,
can be identified through the loop expansion as
$\L_\chi\sim 2\sqrt{2}\pi f_\phi$.
According to the above power counting, 
the 
$\c  O (\epsilon^2)$ 
terms in the (Euclidean) Lagrangian density are
\beq
\c L =\frac{f^2}{8}\Tr(D_{\mu}\Sigma^\dagger D_{\mu}\Sigma)-\lambda\Tr(m_q^{\dagger}\Sigma+m_q\Sigma^\dagger),
\eeq
where the action of the covariant derivative on the coset field is specified by
\beq
D_\mu\Sigma=\partial_\mu\Sigma+ieA_\mu[\mathcal{Q},\Sigma].
\eeq
Higher-order terms encode the short-distance physics;
but, 
these appear at 
$\c O (\epsilon^4)$
in the power counting and will not be required in our calculations.

%%%%%%%%%%%%%%%%%%%%%%%%%%%%%%%%%%%%%
\begin{figure}
\begin{center}
%%%%%%%%%%%%%%%%%%%%%%%%%%%%%%%%%%%%%
\includegraphics[scale=1]{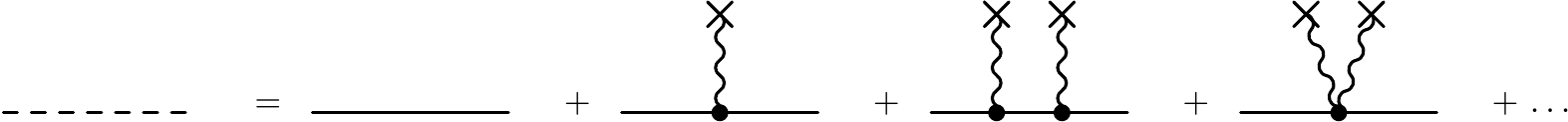}
%%%%%%%%%%%%%%%%%%%%%%%%%%%%%%%%%%%%%
\caption{Charged meson propagator in a large magnetic field. 
Free meson propagators are shown as solid lines; 
external fields are depicted as wiggly lines ending in crosses; 
charge couplings appearing in the $\c O (\epsilon^2)$ chiral Lagrangian density are depicted by filled circles; 
and, 
these must be summed to obtain the propagator (dashed line) in the large-field power counting.}
\label{fig:zero}
%%%%%%%%%%%%%%%%%%%%%%%%%%%%%%%%%%%%%
\end{center}
\end{figure}
%%%%%%%%%%%%%%%%%%%%%%%%%%%%%%%%%%%%%

To study the effects of a large magnetic field on hadrons, 
we choose a uniform magnetic field in the 
$x_3$ direction; 
and, 
for definiteness, 
we implement this field through the choice of gauge 
$A_\mu=(-Bx_2,0,0,0)$. 
According to the power counting assumed in 
Eq.~(\ref{eq:1}), 
the effects of the external magnetic field are non-perturbative with respect to the meson momentum and mass. 
This requires summation of the charge couplings of the Goldstone bosons to the external magnetic field to all orders, 
see Fig.~\ref{fig:zero}. 
In the context of the chiral condensate, 
this summation can be done at the level of the effective action, 
see
Ref.~\cite{Cohen:2007bt}. 
For computation of baryon energy levels, 
we require meson propagators in presence of the magnetic field, 
and these can be determined using Schwinger's proper-time trick%
~\cite{Schwinger:1951nm}.
We utilize Feynman rules in position space throughout, 
for which the propagator of the pseudoscalar meson 
$\phi$ 
having charge 
$Q_\phi$
is given by%
~\cite{Tiburzi:2014zva} 
\beq
G_{\phi}(x,y)
=
e^{i e Q_\phi B \Delta x_{1} \ol x_{2}}
\int_{0}^{\infty}
\frac{ds}{(4\pi s)^{2}}
\frac{e Q_\phi B s}{\sinh(e Q_\phi B s)}e^{-m_{\phi}^{2}s}
\exp\left[
-\frac{e Q_\phi B \, \Delta \vec{x} \, {}_{\perp}^{2}}{4\tanh(e Q_\phi B s)}
-\frac{\Delta x_{3}^{2}+\Delta x_{4}^{2}}{4s}\right],
\label{eq:mesonprop}
\eeq
where the displacement is 
$\Delta x_\mu=x_\mu-y_\mu$, 
the average position is 
$\ol x=\frac{1}{2}(x_\mu+y_\mu)$, 
and the transverse separation squared is 
$\Delta\vec{x} \, {}_{\perp}^{2}=\Delta x_{1}^{2}+\Delta x_{2}^{2}$. 
A few comments regarding the form of the propagator are in order. 
When 
$e Q_\phi B =0$, 
one recovers the Klein-Gordon propagator, 
which has an 
$SO(4)$ symmetry and Euclidean translational invariance in four directions. 
For nonzero values, 
however, 
the integrand in the expression above has only an 
$SO(2)\times SO(2)$ symmetry. 
The phase factor multiplying the integral, 
moreover, 
breaks translational invariance in the $x_2$ direction, 
as well as the 
$SO(2)$ 
symmetry in the plane transverse to the magnetic field. 
The phase factor is gauge dependent; 
consequently, 
the computation of gauge-invariant quantities will reflect 
$SO(2) \times SO(2)$
symmetry and translational invariance.

%%%%%%%%%%%%%%%%
\subsection{Baryon Sector}
%%%%%%%%%%%%%%%%

The na\"ive inclusion of baryons in the chiral Lagrangian introduces a large mass scale which does not vanish in the chiral limit, 
i.e.~$M_B \sim \Lambda_\chi$. 
A systematic way of treating baryons in the chiral Lagrangian is to treat them non-relativistically, 
and the framework of heavy baryon $\chi$PT 
(HB$\chi$PT) 
proves especially convenient%
~\cite{Jenkins:1990jv}. 
In our computation, 
we include the spin-3/2 decuplet degrees of freedom, 
which is necessitated by the three-flavor chiral expansion.%
\footnote{
The basic argument is as follows. 
One cannot justify the computation of $\Sigma$ baryon properties by retaining, 
for example,  
$\pi \Sigma$
and
$K N$ 
loop contributions alone, 
because the 
$K \Delta$
loop contributions represent those from an intermediate-state baryon lying 
\emph{below} 
the 
$\Sigma$. 
Furthermore, 
if one combines the three-flavor chiral limit with the large-$N_c$ limit, 
then both octet and decuplet degrees of freedom are required to produce the correct spin-flavor symmetric loop contributions. 
} 
After the octet baryon mass 
$M_B$ 
is phased away, 
the mass splitting 
$\Delta=M_T - M_B$ 
appears in the decuplet baryon Lagrangian. 
In addition to the chiral power counting in 
Eq.~(\ref{eq:1}), 
we have additionally the HB$\chi$PT power counting
\beq
\frac{k}{M_B}\sim\frac{\Delta}{M_B}\sim\epsilon,
\eeq
where 
$k$ 
is the residual baryon momentum. 
This is the phenomenologically motivated power counting known in the two-flavor case as the small-scale expansion% 
~\cite{Hemmert:1997ye}. 
To leading order, 
which in the baryon sector is 
$\mathcal{O}(\epsilon)$, 
the octet baryon Lagrangian density is given by
\beq
\mathcal{L}
=
- i \, \Tr \left(\ol B \, v \cdot \mathcal{D} B \right)
+
2D \, \Tr \left(\ol B S_\mu\{\mathcal{A}_\mu,B\} \right)
+
2F \, \Tr \left(\ol B S_\mu[\mathcal{A}_\mu,B] \right),
\eeq
where the octet baryons are conventionally embedded in the matrix
\beq
B^i_j=
  \begin{pmatrix}
    \frac{1}{\sqrt{2}}\Sigma^0+\frac{1}{\sqrt{6}}\Lambda & \Sigma^+ & p  \\
    \Sigma^- & -\frac{1}{\sqrt{2}}\Sigma^0+\frac{1}{\sqrt{6}}\Lambda & n  \\
    \Xi^- & \Xi^0 & -\frac{2}{\sqrt{6}}\Lambda
  \end{pmatrix}^i_j
\quad ,\eeq
and the decuplet baryon Lagrangian density to the same order is given by
\beq
\mathcal{L}=\ol T _\mu(-i v \cdot \mathcal{D} +\Delta)T_\mu
+
2\mathcal{H} \, \ol T_\mu \, S \cdot \mathcal{A} \, T_{\mu}
+
2\, \mathcal{C}  \left( \ol T \cdot  \mathcal{A} \, B +\ol B \mathcal{A} \cdot  T
\right).
\eeq
Appearing above is 
$S_\mu$, 
which is the covariant spin operator satisfying the relation
$S_\mu S_\mu=\frac{3}{4}$, 
and
$v_\mu$
is the four-velocity. 
There is the covariant constraint
$v \cdot S = 0$;  
but, 
our computations are restricted to the rest frame
in which 
$v_\mu = (0,0,0,1)$. 
The decuplet baryons are embedded in the completely symmetric flavor tensor,  
$T_{ijk}$, 
in the standard way, 
with the required invariant contractions treated implicitly. 
The coupling of electromagnetism is contained in the vector, 
$\mathcal{V}_\mu$,
and axial-vector, 
$\mathcal{A}_\mu$, 
fields of mesons, 
which have the form
\begin{align}
\mathcal{V}_\mu&=ieA_\mu\mathcal{Q}+\frac{1}{2f^2} \left[\phi, D_\mu\phi \right] + \cdots,
\notag
\\
\mathcal{A}_\mu&=- \frac{1}{f}D_\mu\phi+\cdots
,\end{align}
where ellipses denote terms of higher order than needed for our computation. 
The chirally covariant derivative, 
$\c D_\mu$,  
acts on the octet baryon fields and decuplet baryon fields, respectively as
\begin{align}
(\mathcal{D}_\mu B)^i_j&=\partial_\mu B^i_j+[\mathcal{V}_\mu,B]^i_j
,\notag
\\
(\mathcal{D}_\mu T)_{ijk}&=\partial_\mu T_{ijk}+(\mathcal{V}_\mu)_i^{i'}T_{i'jk}+(\mathcal{V}_\mu)_j^{j'}T_{ij'k}+(\mathcal{V}_\mu)_k^{k'}T_{ijk'}
.\end{align}
Finally the low-energy constants 
$D$, 
$F$, 
and
$\c C$
are chiral-limit values of the axial couplings. 
The decuplet-pion axial coupling 
$\c H$
is not needed in our computations.

As we perform our calculations in position space, 
the static octet baryon propagator required in perturbative diagrams has the form
\beq \label{eq:Bprop}
D_B(x,y)
=
\delta^{(3)}(\vec{x}-\vec{y})
\theta(x_{4}-y_{4}),
\eeq
whereas, the static decuplet propagator is
\beq \label{eq:Tprop}
[D_T(x,y)]_{\mu \nu}
=
\mathcal{P}_{\mu \nu}
\,
\delta^{(3)}(\vec{x}-\vec{y})
\theta(x_{4}-y_{4})
e^{-\Delta(x_{4}-y_{4})}.
\eeq
where the polarization tensor, 
$P_{\mu \nu}$,  
for the spin-3/2 Rarita-Schwinger field is given by 
$\mathcal{P}_{\mu \nu}= \delta_{\mu \nu} - v_\mu v_\nu - \frac{4}{3} S_\mu S_\nu$.

As in the meson sector, 
we partially account for the rather large breaking of 
$SU(3)_V$ 
symmetry. 
For the baryons, 
this is accomplished by treating the baryon mass splittings as their physical values. 
In terms of the octet baryons, 
for example, 
such corrections to baryon masses arise from the following 
$\mathcal{O}(\epsilon^2)$ 
terms in the Lagrangian density
\beq
\label{eq:massops}
\mathcal{L}=b_D\Tr\left(\ol B\,  \{m_q ,B\}\right)+b_F\Tr\left(\ol B\, [m_q,B]\right)+b_\sigma\Tr\left(\ol B \, B\right)\Tr\left(m_q \right),
\eeq
where only the leading such contributions in the chiral expansion are shown. 
The effect of these operators is to lift the degeneracy between the octet baryons. 
Using the physical mass splittings among the various baryons in loop diagrams then corresponds to resummation of the effects of 
Eq.~\eqref{eq:massops}, 
and analogously those for the decuplet fields, 
into the propagators. 
Accordingly the propagators in 
Eqs.~\eqref{eq:Bprop} and \eqref{eq:Tprop} 
are modified away from their 
$SU(3)_V$ 
symmetric forms. 
As we work in the strong isospin limit, 
isospin-averaged baryon mass splittings are utilized in each of these propagators.

%%%%%%%%%%%%%%%%%%%%%%%%%%%%%%%%%%%%
\section{Determination of Octet Baryon Energies in Magnetic Fields}   %
%%%%%%%%%%%%%%%%%%%%%%%%%%%%%%%%%%%%
\label{s:energy}

Having spelled out the required elements of chiral perturbation theory in both the meson and baryon sectors, 
we now proceed to compute the energy levels of the octet baryons in large magnetic fields. 
There are both tree-level and loop contributions, 
and we compute the octet baryon self energies to 
$\mathcal{O}(\epsilon^3)$
in the combined chiral and heavy baryon expansion.  
Notice that for the loop contributions, 
the power counting, 
Eq.~(\ref{eq:1}),  
dictates that charged-meson propagators include the magnetic field non-perturbatively compared to the meson mass and momentum.  
Baryon propagators, 
by contrast, 
are affected by the external magnetic field perturbatively.

%%%%%%%%%%%%%%%%%%%%%%%%%%%
\subsection{Tree-Level Contributions}

The tree-level contributions to the energies are simplest and therefore handled first. 
Local operators contribute to the energies at 
$\c O(\epsilon^2)$. 
These are the octet baryon magnetic moment operators 
\beq\label{eq:colgla}
\mathcal{L}
=
\frac{e}{2M_N}
\left[
\mu_D
\Tr \left(\ol B \, S_{\mu\nu} \{\mathcal{Q},B\} \right)
+
\mu_F
\Tr \left(\ol B \, S_{\mu\nu} [\mathcal{Q},B] \right) \right] F_{\mu\nu},
\eeq
where we have made the abbreviation, 
$S_{\mu\nu}= \epsilon_{\mu \nu \a \b} \, v_\a S_\b$.
The low-energy constants are the Coleman-Glashow magnetic moments, 
$\mu_D$
and
$\mu_F$%
~\cite{Coleman:1961jn}. 
The remaining 
$\c O(\epsilon^2)$
contribution to the octet baryon energies arises from the kinetic-energy term of the Lagrangian density. 
In the heavy baryon formulation, 
this term is given by
\beq
\mathcal{L}=- \Tr \left(\ol B \frac{\mathcal{D}^2_\perp}{2M_B} B \right), 
\eeq
where $(\mathcal{D_\perp})_\mu=\mathcal{D}_\mu - v_\mu(v \cdot \mathcal{D})$, 
and the coefficient of this operator is exactly fixed to unity by reparametrization invariance%
~\cite{Luke:1992cs}. 
For neutral baryons, 
this contribution vanishes for states at rest. 
For baryons of charge 
$Q$, 
the gauged kinetic term produces eigenstates that are Landau levels,  
which,  
for zero longitudinal momentum,  
$k_3 = 0$, 
the energy eigenvalues are given by 
\beq
E_{n_L}
=
\frac{|QeB|}{M_B}\left(n_L+\frac{1}{2}\right). 
\eeq
In order to maintain the validity of the power counting, 
we are necessarily restricted to the lower Landau levels characterized by parametrically small values of the quantum number
$n_L$. 
We restrict our analysis below to the lowest Landau level,
$n_L=0$. 
While the Landau levels depend non-perturbatively on the magnetic field, 
the Landau levels of intermediate-state baryons affect energy levels at 
$\c O(\epsilon^4)$;
and, fortunately can be dropped in our calculation.

%%%%%%%%%%%%%%%%%%%%%%%%%%%%%%%%%%%%%%%
\begin{figure}
\begin{center}
%%%%%%%%%%%%%%%%%%%%%%%%%%%%%%%%%%%%%%%
\includegraphics[scale=1]{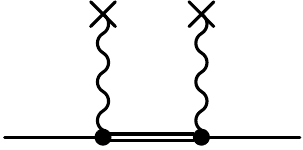}
%%%%%%%%%%%%%%%%%%%%%%%%%%%%%%%%%%%%%%%
\caption{Decuplet pole diagram which produces 
$\c O(\epsilon^3)$ 
contributions to the octet baryon energies.
The octet baryons are shown with solid lines, 
while decuplet baryons are shown with double lines. 
External fields are shown with wiggly lines terminating in crosses. 
}
\label{fig:two}
%%%%%%%%%%%%%%%%%%%%%%%%%%%%%%%%%%%%%%%
\end{center}
\end{figure}
%%%%%%%%%%%%%%%%%%%%%%%%%%%%%%%%%%%%%%%

Another operator that enters at order 
$\c O(\epsilon^2)$
is the magnetic dipole transition operator between the decuplet and octet baryons, 
which takes the form%
~\cite{Butler:1992pn}
\beq \label{eq:trans}
\mathcal{L}
=
\mu_{U}
\,
\sqrt{\frac{3}{2}} 
\frac{ie}{M_N}
\left(\ol B \, S_\mu \mathcal{Q} \, T_\nu + \ol T_{\mu}\mathcal{Q} \, S_{\nu}B \right)
F_{\mu\nu},
\eeq
where the $U$-spin \cite{Lipkin:1964zza} symmetric transition moment,
$\mu_U$, 
can be determined from the measured electromagnetic decay widths of decuplet baryons. 
The dipole transition operator contributes to octet baryon energies at 
$\c O(\epsilon^3)$
through two insertions in the tree-level diagram shown in 
Fig.~\ref{fig:two}. 
Essentially the addition of a uniform magnetic field leads to mixing between the octet and decuplet baryons via 
Eq.~\eqref{eq:trans}, 
and the decuplet pole diagram represents the first perturbative contribution from this mixing.  
The large size of the transition moment, 
$\mu_U$, 
is a well-known issue in the description of nucleon magnetic polarizabilities,
for an early investigation of the delta-pole contribution, see \cite{Mukhopadhyay:1993zx}. 
Consequently the 
$\c O(B^2)$
contribution to the octet baryon energies in magnetic fields will be problematic;
and, 
we investigate three scenarios for this contribution in Sec.~\ref{s:scenarios} below.

%%%%%%%%%%%%%%%%%%%%%%%%%%%%%%%%%%%%%%%
\begin{table}
\def\arraystretch{1.5}
\caption{$U$-spin symmetric coefficients for tree-level contributions to the octet baryon energies appearing in Eq.~\eqref{eq:tree}.} 
\label{tab:one}
\medskip
\begin{tabular}{|c|rrr|}
\hline
$B$ 
&
$\quad Q$
&
$\quad \alpha_D$
& 
$\quad \alpha_T$
\tabularnewline
\hline 
$p$, 
$\Sigma^+$ 
&
1
& 
$\frac{1}{3}$
&
$\frac{1}{3}$
\tabularnewline
$n$, 
$\Xi^0$
&
$0$
&
$- \frac{2}{3}$
&
$\frac{1}{3}$
\tabularnewline
$\Lambda$
&
$0$
& 
$- \frac{1}{3}$
& 
$\frac{1}{4}$
\tabularnewline
$\Sigma^{0}$
& 
$0$
&
$\frac{1}{3}$
&
$\frac{1}{12}$
\tabularnewline
$\Sigma^{0}\to\Lambda$
&
$0$
&
$\frac{1}{\sqrt{3}}$
&
$-\frac{1}{4\sqrt{3}}$
\tabularnewline
$\Sigma^-$, 
$\Xi^{-}$
&
$-1$
&
$\frac{1}{3}$
&
$0$
\tabularnewline
\hline
\end{tabular}
\end{table}
%%%%%%%%%%%%%%%%%%%%%%%%%%%%%%%%%%%%%%%

Considering all tree-level contributions, 
the resulting energy levels of the octet baryons to 
$\c O(\epsilon^3)$
are given by
\beq \label{eq:tree}
E^\text{tr}
=
M_B
+
\frac{|QeB|}{2M_B}
-
\left(
\alpha_D
\mu_D
+
Q
\mu_F
\right)
\frac{eB \, \sigma_3}{2 M_N}
-
\frac{\alpha_T \, \mu_U^2}{\Delta}\left(\frac{eB}{2M_N}\right)^2,
\eeq
where
neutral particles are taken at rest, 
and charged particles are taken in their lowest Landau level with zero longitudinal momentum. 
The $U$-spin symmetric coefficients are labeled by 
$Q$, 
$\alpha_D$, 
and 
$\alpha_T$. 
These coefficients depend on the octet baryon state of interest,
and are given in 
Table~\ref{tab:one}. 
Notice that the octet magnetic moment operators lead to a Zeeman effect, 
with the energies depending on the projection of spin along the magnetic field axis, 
$\sigma_3$.

%%%%%%%%%%%%%%%%%%%%%%%%%%%%%%%%%%%%%%%
\subsection{Meson-Loop Contributions}
%%%%%%%%%%%%%%%%%%%%%%%%%%%%%%%%%%%%%%%

%%%%%%%%%%%%%%%%%%%%%%%%%%%%%%%%%%%%%%%
\begin{figure}
\begin{center}
\medskip
%%%%%%%%%%%%%%%%%%%%%%%%%%%%%%%%%%%%%%%
\includegraphics[scale=1]{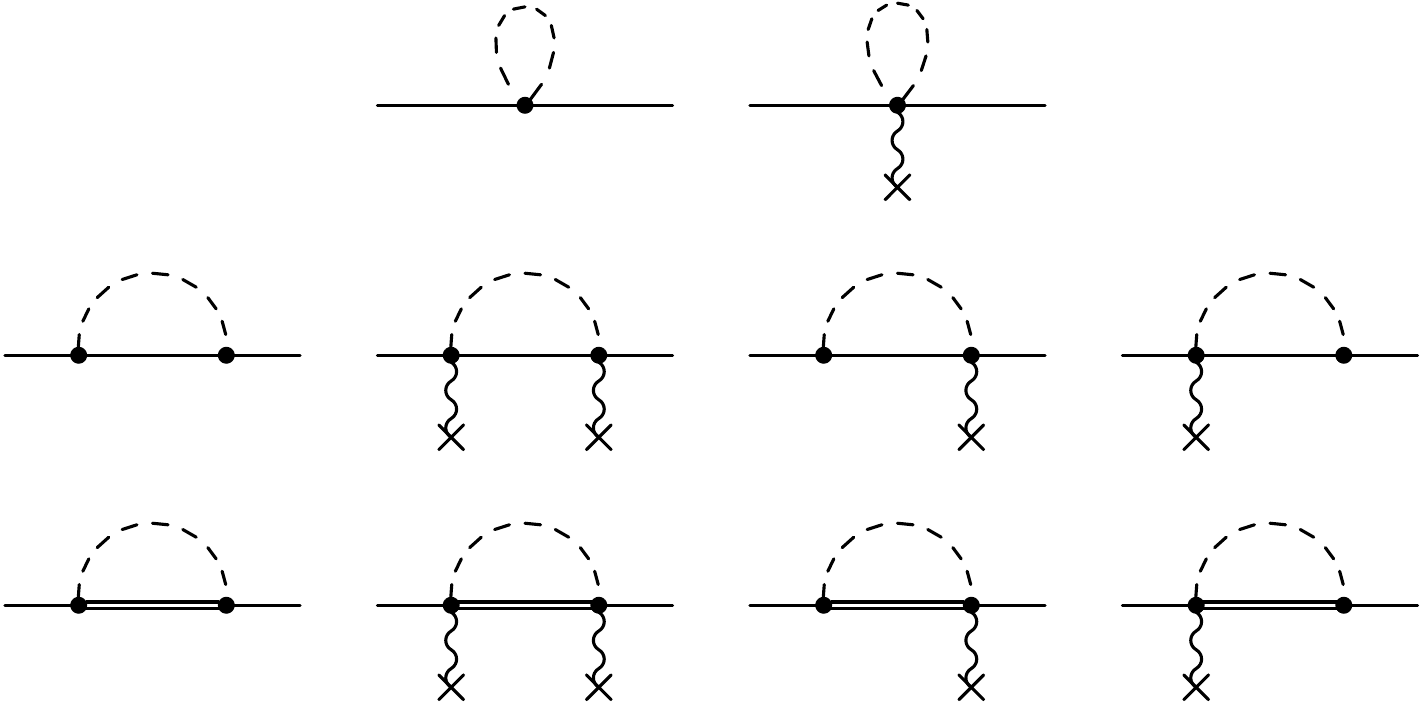}
%%%%%%%%%%%%%%%%%%%%%%%%%%%%%%%%%%%%%%%
\caption{
Loop contributions to the octet baryon energies at 
$\c O(\epsilon^3)$. 
The single lines represent octet baryons; 
the double lines represent decuplet baryons; 
and, 
wiggly lines represent the external magnetic field. 
The dashed lines represent mesons propagating in the magnetic field, 
see the diagrammatic depiction in Fig.~\ref{fig:zero}. 
While generated from couplings in the Lagrangian density, 
diagrams in the first row identically vanish. 
}
\label{fig:one}
%%%%%%%%%%%%%%%%%%%%%%%%%%%%%%%%%%%%%%%
\end{center}
\end{figure}
%%%%%%%%%%%%%%%%%%%%%%%%%%%%%%%%%%%%%%%

Beyond trees, 
meson loops contribute to the baryon energies and such contributions are non-analytic with respect to the meson mass and magnetic field.
The diagrams which contribute at 
$\mathcal{O}(\epsilon^3)$
are depicted in 
Fig.~(\ref{fig:one}). 
The meson tadpole diagrams vanish, 
either by virtue of time-reversal invariance or by the gauge condition, 
$v \cdot A = 0$. 
The two sets of four sunset diagrams shown are connected by gauge invariance. 
There is one set for intermediate-state octet baryons and another set for intermediate-state decuplet baryons. 
A set of four sunset diagrams is best expressed as a single sunset, 
arising from a gauge covariant derivative at each meson-baryon vertex.

It is useful to sketch the required position-space computation of the loop contributions to baryon energies in our approach. 
Each loop contribution contains a product of charge and
Clebsch-Gordan coefficients, 
along with other numerical factors. 
Putting aside such factors 
for simplicity, 
the amputated contribution to the two-point function of the octet baryon $B$,
denoted 
$\delta D_B(x',x)$,
in the case of an intermediate-state octet baryon, 
$B'$, 
is given by 
\beq
\delta D_B(x',x)
=
\sigma_i 
D_{B'}(x',x)
\sigma_j 
D'_i D_j
G_\phi(x',x),
\label{eq:pertB}
\eeq
whereas, in the case of intermediate-state decuplet baryons, 
$T$,  
the corresponding contribution is of the form
\beq
\delta D_B(x',x)
=
[D_T(x',x)]_{ij}
D'_i D_j
G_\phi(x',x).
\label{eq:pertT}
\eeq
Above, the primed gauge covariant derivative depends on the coordinate 
$x'$, 
which appears both in the partial derivative and gauge potential. 
Perturbative corrections to the octet baryon energies, 
$\delta E$,  
are identified by projecting the amputated two-point function onto vanishing residual baryon energy, 
$k_4=0$. 
The term linear in 
$k_4$
produces the wave-function renormalization, 
however, 
this contributes to baryon energies at 
$\c O(\epsilon^4)$, 
which is beyond our consideration. 
Putting the baryon $B$ on-shell, 
we have
\beq
\int_{-\infty}^\infty d(x'_4-x_4)
\, \delta D_B(x',x)
=
-\delta^3(\vec{x}'-\vec{x})
\, 
\delta E
,\eeq
where the delta-function arises from translational invariance, 
which is expected because the breaking of translational invariance is a gauge artifact. 
The loop correction to the baryon energy, 
$\delta E$,
is conveniently decomposed into spin-dependent and spin-independent contributions, 
in the form
\begin{equation}
\label{eq:loopy}
\d E 
= 
- e B \sigma_3 \, \d E_1 + \d E_2
.\end{equation}

%%%%%%%%%%%%%%%%%%%%%%%%%%%%%
%%%%%%%%%%%%%%%%%%%%%%%%%%%%%
\begin{table}[h!]
\begin{center}
\def\arraystretch{1.5}
\caption{Coefficients 
$\c A_\c B$ 
and mass splittings
$\D_\c B$, 
for intermediate-state baryons
$\c B$.
The entries are grouped according to the external octet states 
$B$, 
except for the transition 
$\Sigma^0 \to \Lambda$, 
which represents contributions to the off-diagonal matrix element. 
For each state $B$, 
factors for the contributing intermediate-state baryons are listed, 
along with the corresponding charged loop meson, 
the quantum numbers of which are fixed by flavor conservation. 
The intermediate-state baryons are both octet, 
for which $\c B = B'$, 
and 
decuplet, 
for which $\c B = T$.
If an intermediate-state baryon is not listed, 
its contribution vanishes. 
Notice that in the case of intermediate-state decuplet baryons, 
all coefficients 
$\c A_T$
are proportional to the baryon-transition axial coupling squared, 
$\c C^2$. 
} 
\label{tab:two}
\medskip
%%%%%%%%%%%%%%%%%%%%%%%%%%%%%
\begin{tabular}{ |L||L|C||L|C|C||L|C|C||}
 \hline
 \hline
B \qquad \qquad \qquad
& \phi \qquad \quad
& \quad Q_{\phi} \quad \quad
& B' \qquad \quad 
& \mathcal{A}_{B'} 
& \Delta_{B'}
& T \qquad \qquad \quad
&  \quad \c A_T /\c C^2 \quad \quad
& \Delta_{T} 
\tabularnewline
\hline
\hline 
 p & \pi & + 1 & n  & (D+F)^{2} &  0 & \Delta^{0} &\frac{1}{3} & M_\Delta-M_N \\
    &      & - 1 &  &   &  & \Delta^{++} &1 &M_\Delta-M_N \\
\cline{2-9}
 & K & +1 &  \Lambda  & \frac{1}{6}(D+3F)^{2}& M_\Lambda-M_N  & & &  \\
 &  & +1  &  \Sigma^{0}  & \frac{1}{2}(D-F)^{2}& M_\Sigma-M_N & \Sigma^{*0} &\frac{1}{6} &M_{\Sigma^{*}}-M_N \\
\hline
\hline
n   & \pi & + 1 & & &  & \Delta^{-} &1 & M_\Delta-M_N  \\
     &  & -1 & p  & (D+F)^{2}&  0  & \Delta^{+} & \frac{1}{3} &M_\Delta-M_N \\
\cline{2-9}
  &K & +1 &  \Sigma^{-}  & (D-F)^{2}& M_\Sigma-M_N  & \Sigma^{*-} &\frac{1}{3} &M_{\Sigma^{*}}-M_N \\
\hline
\hline
\Lambda &\pi & +1 &  \Sigma^{-}  & \frac{2}{3}D^{2} &M_\Sigma-M_\Lambda &  \Sigma^{*-} & \frac{1}{2} & M_{\Sigma^{*}}-M_\Lambda \\
 & & -1 &  \Sigma^{+}  & \frac{2}{3}D^{2} & M_\Sigma-M_\Lambda& \Sigma^{*+} & \frac{1}{2} &M_{\Sigma^{*}}-M_\Lambda\\
\cline{2-9}
  & K & +1 & \Xi^{-}  & \frac{1}{6}(D-3F)^{2} & M_\Xi-M_\Lambda  &  \Xi^{*-} &\frac{1}{2} & M_{\Xi^{*}}-M_\Lambda \\  
  & &  -1 & p  & \frac{1}{6}(D+3F)^{2} & M_N-M_\Lambda  &  &    &  \\    
\hline
\hline
\Sigma^{+} & \pi & +1  &  \Lambda  & \frac{2}{3}D^{2}& M_\Lambda-M_\Sigma & & & \\
& & +1  &  \Sigma^{0} & 2F^{2}& 0& \Sigma^{*0} &\frac{1}{6} & M_{\Sigma^{*}}-M_\Sigma \\
\cline{2-9}
&K & +1  &  \Xi^{0} & (D+F)^{2} &M_\Xi-M_\Sigma&  \Xi^{*0} &\frac{1}{3} & M_{\Xi^{*}}-M_\Sigma \\
& & -1 & & &   &  \Delta^{++} &1 & M_\Delta-M_\Sigma \\
\hline
\hline
\Sigma^{0} &\pi & +1  &  \Sigma^{-}  & 2F^{2} & 0 &  \Sigma^{*-} &\frac{1}{6} &M_{\Sigma^{*}}-M_\Sigma\\
& & -1 &  \Sigma^{+}  & 2F^{2} & 0 & \Sigma^{*+} &\frac{1}{6} &M_{\Sigma^{*}}-M_\Sigma\\
\cline{2-9}
& K & + 1&  \Xi^{-}  & \frac{1}{2}(D+F)^{2} & M_\Xi-M_\Sigma  &  \Xi^{*-} & \frac{1}{6} & M_{\Xi^{*}}-M_\Sigma\\  
& & -1 & p & \frac{1}{2}(D-F)^{2} & M_N-M_\Sigma  &  \Delta^{+} & \frac{2}{3} & M_\Delta-M_\Sigma\\    
\hline
\hline
\Sigma^{-} & \pi & -1 &  \Lambda  &\frac{2}{3}D^{2} & M_\Lambda-M_\Sigma & & & \\
& & -1 &  \Sigma^{0}  & 2F^{2}& 0& \Sigma^{*0} &\frac{1}{6} &M_{\Sigma^{*}}-M_\Sigma\\
\cline{2-9}
& K & -1  &  n  & (D-F)^{2} & M_N-M_\Sigma & \Delta^{0} & \frac{1}{3} &M_\Delta-M_\Sigma \\
\hline
\hline
\Xi^{0} & \pi & +1 &  \Xi^{-}  & (D-F)^{2} & 0 &   \Xi^{*-} &\frac{1}{3} &M_{\Xi^{*}}-M_\Xi \\
\cline{2-9}
&K & +1 &&& & \Omega^{-} & 1 & M_\Omega-M_\Xi \\
 & & -1 &  \Sigma^{+}  & (D+F)^{2} & M_\Sigma-M_\Xi & \Sigma^{*+} &\frac{1}{3} &M_{\Sigma^{*}}-M_\Xi \\
\hline
\hline
\Xi^{-} & \pi & - 1&  \Xi^{0} &  (D-F)^{2} & 0 & \Xi^{*0}& \frac{1}{3} &M_{\Xi^{*}}-M_\Xi \\
\cline{2-9}
& K & - 1 & \Lambda & \frac{1}{6}(D-3F)^{2} & M_\Lambda-M_\Xi & & &  \\
& & -1 & \Sigma^{0} & \frac{1}{2}(D+F)^{2} & M_\Sigma-M_\Xi& \Sigma^{*0} &\frac{1}{6} & M_{\Sigma^{*}}-M_\Xi \\
\hline
\hline
\Sigma^{0}\to\Lambda & \pi & +1 &  \Sigma^{-} & \phantom{-} \frac{2}{\sqrt{3}}DF & 0 &  \Sigma^{*-} &-\frac{1}{2 \sqrt{3}} &M_{\Sigma^{*}}-M_\Sigma \\
  & & -1 & \Sigma^{+} & -\frac{2}{\sqrt{3}}DF & 0 & \Sigma^{*+}&\phantom{-} \frac{1}{2 \sqrt{3}} &M_{\Sigma^{*}}-M_\Sigma \\
\cline{2-9}
  &K & + 1 &  \Xi^{-}  & - \frac{1}{2 \sqrt{3}} (D+F)(D-3F) & M_\Xi-M_\Sigma &  \Xi^{*-} &- \frac{1}{2 \sqrt{3}} & M_{\Xi^{*}}-M_\Sigma \\  
  & & -1 &  p & - \frac{1}{2 \sqrt{3}} (D-F)(D+3F) & M_N-M_\Sigma  & &  & \\    
\hline
\hline
\end{tabular}
\end{center}
\end{table}

Careful computation of the gauge-covariant derivatives acting on the meson propagator, 
Eq.~\eqref{eq:mesonprop},
contraction of the vector indices, 
and subsequent spin algebra produces the amputated contributions to the two-point function required in 
Eqs.~\eqref{eq:pertB} and \eqref{eq:pertT}. 
Carrying out the integral over the relative time
and
appending the Clebsch-Gordan coefficients, 
along with other numerical factors, 
leads to the spin-dependent and spin-independent loop contributions%
~\cite{Tiburzi:2008ma},
which are given by
\begin{align*}\label{eq:main}
\delta E_{1}
&=
\sum_{\c B}
\c A_\c B \,
\c S_{1\c B} \,
\frac{Q_{\phi} m_\phi}{(4 \pi f_\phi)^2}
\, 
F_1 \left( \frac{|eB|}{m_\phi^2}, \frac{\Delta_\c B}{m_\phi} \right)
,\\
\delta E_{2}
&=
\sum_{\c B}
\c A_\c B \,
\c S_{2 \c B}
\frac{m_\phi^3}{(4 \pi f_\phi)^2}
\,
F_2 \left( \frac{|eB|}{m_\phi^2}, \frac{\Delta_\c B}{m_\phi} \right)
.\numberthis
\end{align*}
These expressions have been written using a compact notation. 
Firstly, 
the external-state baryon 
$B$
has been treated implicitly to avoid an accumulation of labels. 
Each contribution to the energy features a sum over the contributing loop baryons, 
$\c B$, 
which are either octet baryons, 
$B'$, 
or decuplet baryons, 
$T$. 
Given the external state 
$B$, 
and internal baryon 
$\c B$,
the corresponding loop meson 
$\phi$
is uniquely determined; 
hence, 
we do not additionally sum over the charged meson states
$\phi$.  
Products of Clebsch-Gordan coefficients and axial couplings are defined to be
$\c A_\c B$,
and these appear in 
Table~\ref{tab:two}. 
The multiplicative 
$\c S$
factors arise from the spin algebra. 
For the spin-dependent contributions, 
we have 
$\c S_{1\c B}$, 
which takes the values
$\c S_{1B'} = 1$
and
$\c S_{1T} = - \frac{1}{3}$, 
for all
$\c B = B'$
and
$\c B = T$, 
respectively. 
For the spin-independent contributions, 
we have the factor 
$\c S_{2\c B}$, 
which takes the values
$\c S_{2B'} = 1$
and
$\c S_{2T} = \frac{2}{3}$,
which also depend only on the spin of the intermediate-state baryon. 
The arguments of loop functions depend on the baryon mass splitting, 
denoted by 
$\D_\c B$, 
which is defined as 
$\D_\c B = M_\c B - M_B$.
For clarity, 
the splittings are also provided in 
Table~\ref{tab:two}.  
Finally, 
the loop functions, 
$F_1(x,y)$
and
$F_2(x,y)$,
can be written in terms of integrals over the proper-time, 
for which the $x$- and $y$-dependence factorizes in the integrands. 
For the spin-dependent loop contributions, 
we have
\beq \label{eq:F1}
F_1(x,y)
=
\int_0^\infty 
ds \, 
f(x,s) \, g(y,s)
,\eeq
where the magnetic field dependence enters through the function 
\beq
f(x,s)
=
\frac{\pi^{1/2}}{s^{3/2}}~\left(\frac{x s}{\sinh(x s)}-1\right)
.\eeq
Notice that this function is even with respect to 
$x$. 
For this reason, 
we need not include the charge of the loop meson in the argument, 
because 
$|Q_\phi| = 1$. 
The remaining function, 
$g(y,s)$,
encodes the dependence on the baryon mass splitting, 
and is given by
\beq
g(y,s)
=
e^{- s (1-y^2)}
\text{Erfc}\left( y \sqrt{s} \, \right)
.\eeq
The loop function 
$F_2(x,y)$, 
which is relevant for spin-independent contributions to the energy,
can be written in terms of the same auxiliary functions
\beq \label{eq:F2}
F_2(x,y)
=
-
\int_0^\infty ds \,
f(x,s) \frac{d}{ds} g(y,s)
.\eeq
Notice that both loop functions vanish in vanishing magnetic fields, 
which is a consequence of 
$f(0,s) = 0$. 
For the spin-independent contributions to the energy, 
this vanishing implies that all chiral corrections to the baryon mass have been renormalized into the physical value of 
$M_B$. 
For the spin-dependent contributions to the energy, 
the vanishing implies that chiral corrections to the baryon magnetic moments have been renormalized into the physical magnetic moments. 
For completeness, 
the 
$\chi$PT
corrections to magnetic moments in our approach are provided in Appendix~\ref{s:magpol}.

%%%%%%%%%%%%%%%%%%%%%%%
\subsubsection*{Behavior of Loop Functions}%
%%%%%%%%%%%%%%%%%%%%%%%

\begin{figure}[t]
%%%%%%%%%%%%%%%%
\resizebox{\linewidth}{!}{
\includegraphics[height=5cm]{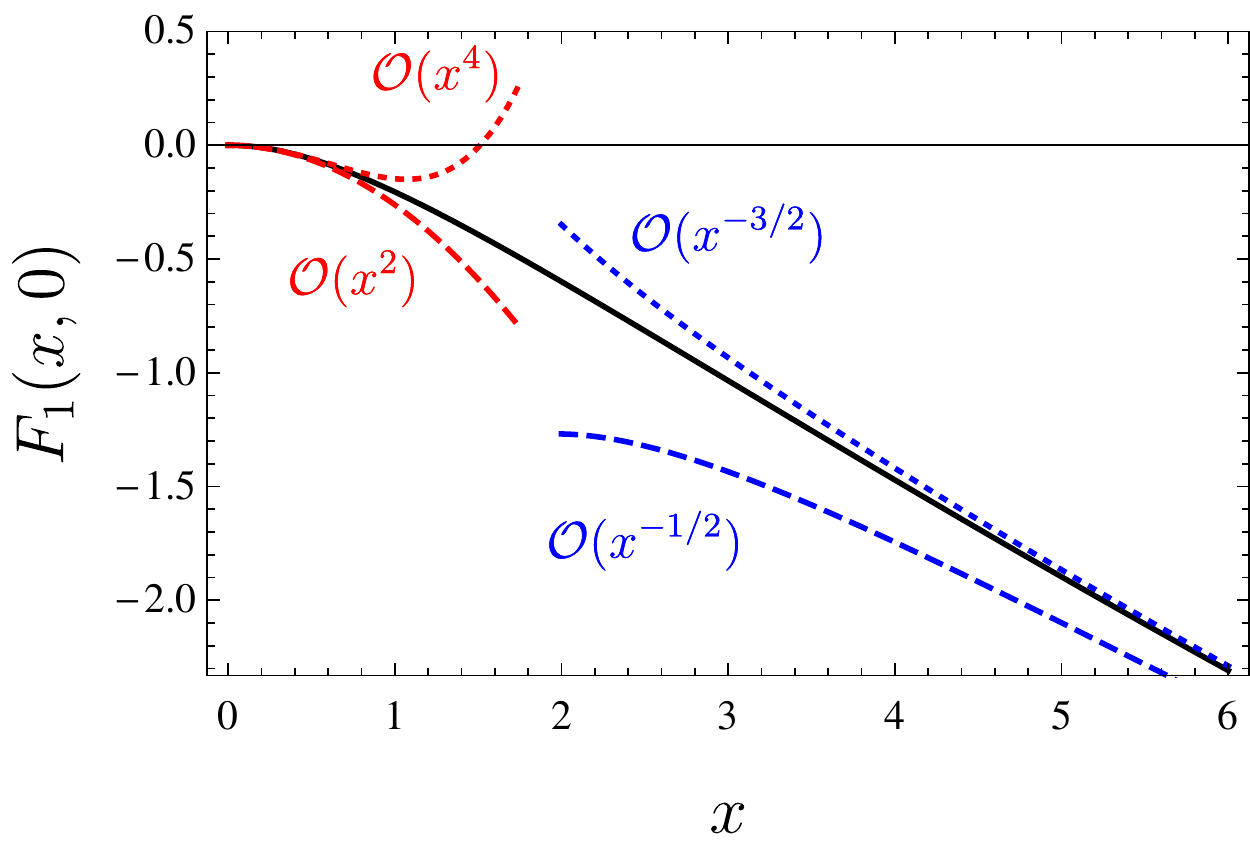}
$\qquad$
\includegraphics[height=5cm]{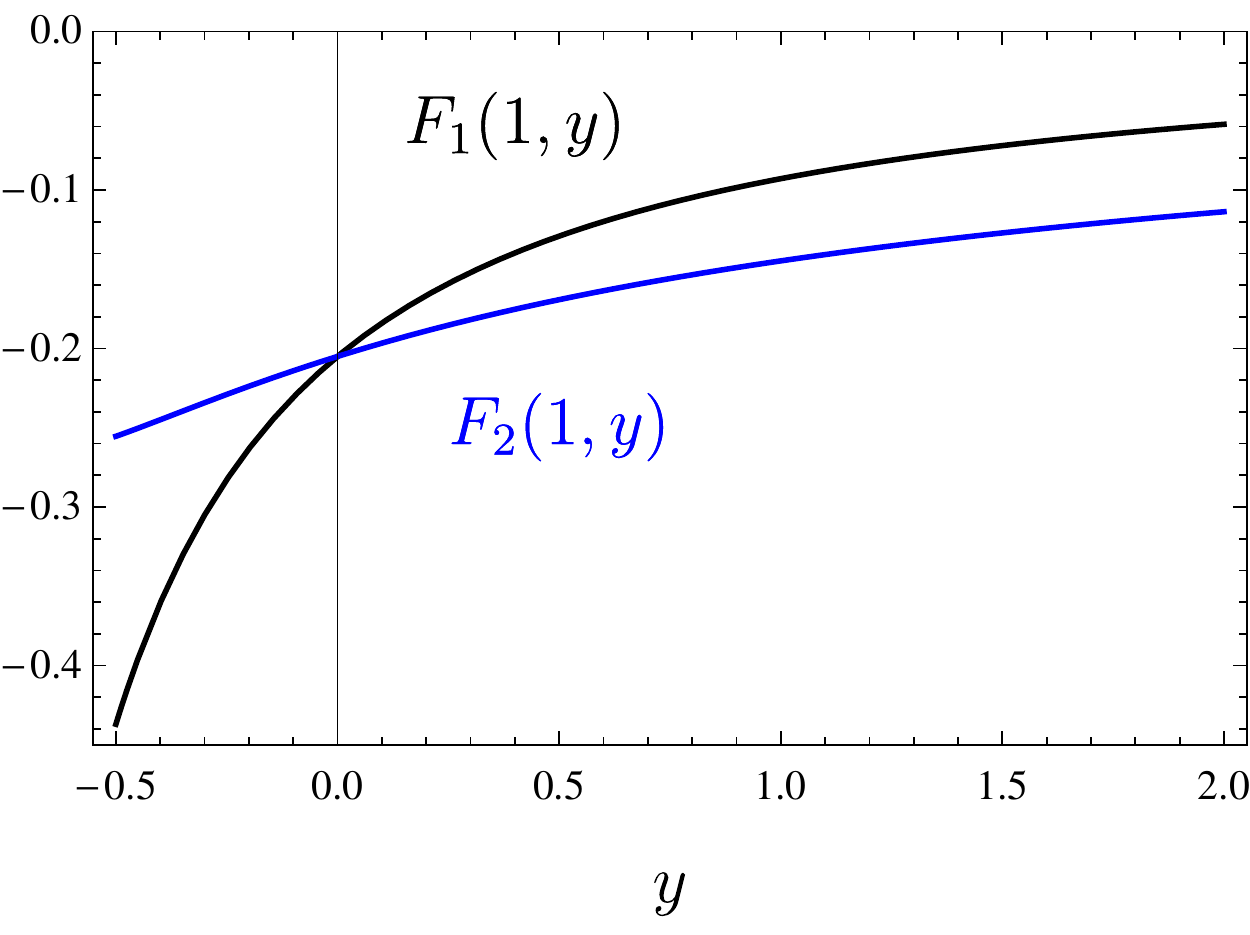}
}
%%%%%%%%%%%%%%%%
\caption{
Behavior of the non-analytic loop functions
$F_1(x,y)$
and
$F_2(x,y)$
appearing in 
Eqs.~\eqref{eq:F1} and \eqref{eq:F2}, 
respectively. 
The left panel shows the $x$-dependence of 
$F_1(x,0)$
(solid), 
along with the small- and large-$x$ asymptotic behavior
(dashed and dotted).
Asymptotic approximations are labeled by the order to which they are valid. 
Notice that the small-$x$ expansions have been plotted beyond their range of applicability.
The right panel shows the $y$-dependence of each function 
evaluated at 
$x = 1$, 
which corresponds to a magnetic field strength satisfying 
$| e B | = m_\phi^2$. 
The range of 
$y$ 
values plotted encompasses the values required by the intermediate-state baryons in loop diagrams. 
Notice that the functions become identical for 
$y = 0$. 
}
\label{fig:dif7}
\end{figure}
%%%%%%%%%%%%%%%%
%%%%%%%%%%%%%%%%

Before presenting the results for octet baryon energies in large magnetic fields, 
it is instructive to consider the general behavior of the non-analytic loop functions.  
The octet baryon energies contain sums over these loop functions evaluated for various values of the mass parameters, 
see Eq.~(\ref{eq:main}). 
To exhibit the general behavior,
we compare the loop functions 
$F_1$
and
$F_2$
with their small- and large-$x$ asymptotic behavior, 
over a range of 
$y$
values required by the intermediate-state baryons.

The small-$x$ behavior is relevant for perturbatively weak magnetic fields, 
with the first non-vanishing term occurring at order 
$x^2$. 
In the case of 
$F_2$, 
the 
$\c O(x^2)$ 
term is a contribution to the effect on the energy from the magnetic polarizability. 
While the small-$x$ expansion can be carried out for general values of 
$y > -1$, 
we cite only the simple expression for 
$y=0$. 
Notice that for this particular value, 
we have
$F_1(x,0) = F_2(x,0)$, 
and further that
\beq
F_1(x,0) 
&\overset{x\ll1}{=}&
- \frac{\pi}{12} x^2
\left[ 
1
- 
\frac{7}{16} 
x^2
+ 
\c O(x^4)
\right]
.\eeq
The large-$x$ behavior, 
by contrast,  
becomes relevant in the chiral limit. 
This limit, 
furthermore, 
can only be taken provided 
$y \geq 0$.%
\footnote{
When 
$\D_\c B < 0$, 
the corresponding value of 
$y$
approaches
$y \to - \infty$
in the chiral limit. 
The loop functions 
$F_1(x,y)$
and
$F_2(x,y)$
themselves
become infinite in this limit, 
and the corresponding baryon states no longer exist in the low-energy spectrum. 
}
The contributions with 
$y > 0$, 
vanish when the chiral limit is taken, 
and these correspond to intermediate states that decouple. 
The simplest expression arises for 
$y = 0$, 
for which the different loop functions are the same; 
and,  
we have
\beq
F_1(x,0)
\overset{x\gg1}{=}
2 \pi \sqrt{x}
\left[
( 1  - 2^{-\frac{1}{2}})
\zeta_{\frac{1}{2}}
+ 
\frac{1}{\sqrt{x}}
- 
\frac{1 - 2^{- \frac{3}{2}}}{ 2 x}
\zeta_{\frac{3}{2}}
+
3 \frac{1-2^{-\frac{5}{2}}}{8 x^2}
\zeta_{\frac{5}{2}}
+ 
\c O(x^{-3})
\right]
,\eeq
where
$\zeta_z$
is used to denote the 
Riemann
zeta-function, 
$\zeta(z)$. 
Note that the fractional power of $x$ appearing within the brackets is the only such term in the asymptotic series.% 
\footnote{
Due to the dominant
$\sqrt{x}$ 
factor exhibited in the chiral limit,
we have from 
Eq.~\eqref{eq:main} 
the behavior of the loop contributions:
$\d E_1 \sim | e B |^{1/2}$,
and
$\d E_2 \sim m_\phi^2 \, |e B|^{1/2}$. 
As a result, 
only the spin-dependent loop contributions survive; 
and, 
by virtue of 
Eq.~\eqref{eq:loopy}, 
we have the chiral-limit behavior
$\delta E \sim - e B \sigma_3  | e B|^{1/2}$,
which deviates from a linear Zeeman effect. 
}

In Fig.~\ref{fig:dif7}, 
the behavior of the loop function 
$F_1(x,0) = F_2 (x,0)$
is shown as a function of 
$x$. 
Additionally shown are the small- and large-$x$ 
asymptotic limits, 
with the function interpolating between these extremes. 
The figure, 
moreover, 
illustrates the dependence on 
$y$
(for the particular value $x = 1$)
by showing the loop functions over a range spanning the smallest and largest values of 
$y$
required by the intermediate-state baryons.
The $\Lambda$ contribution to the $\Sigma$ energies requires the smallest value, 
$y \approx - \frac{1}{2}$;
while,
the delta contribution to the nucleon energies requires the largest value, 
$y \approx 2$. 
The importance of the loop functions generally increases with decreasing values of 
$y$, 
which is physically reasonable because lower-lying states have smaller $y$ values and should give more important non-analytic contributions.

%%%%%%%%%%%%%%%%%%%%%%%%
\subsection{Complete Third-Order Calculation}%
%%%%%%%%%%%%%%%%%%%%%%%%
\label{s:third}

Accounting for the tree-level and loop contributions, 
as well as the renormalization in vanishing magnetic fields, 
we have the general expression for the octet baryon energies valid to 
$\c O(\epsilon^3)$
\beq
E
=
M_B
+
\frac{|QeB|}{2M_B}
-eB\sigma_3
\left(\frac{\mu_B}{2M_N}+\delta E_1 \right)
-\frac{\a_T \, \mu_U^2 }{\Delta_T}
\left(\frac{eB}{2M_N}\right)^2+\delta E_2.
\label{eq:energies}
\eeq
In the above expression, 
neutral baryons are taken at rest, 
while charged baryons are in their lowest Landau level with vanishing longitudinal momentum. 
The baryon energies depend on known parameters:
the hadron masses, baryon magnetic moments, and meson decay constants. 
The axial couplings are reasonably well constrained from phenomenological analyses,
and we adopt the values 
$D=0.61$,
$F=0.40$,
$\c C=1.2$%
~\cite{Butler:1992ci}.
The transition dipole moment, 
$\mu_U$, 
will be discussed below in conjunction with the nucleon magnetic polarizabilities. 
For reasons that will become clear, 
we do not attempt to propagate uncertainties on parameters or from neglected higher-order contributions.

While the above expression applies equally well to all members of the baryon octet, 
there is the additional feature of mixing between the 
$\S^0$
and
$\L$
baryons.
Because coupling to the external magnetic field breaks isospin symmetry
(but preserves $I_3$), 
mixing is possible between these two 
$I_3 = 0$ 
baryon states. 
In this two-state system, 
we must consider the magnetic-field dependent energy matrix
\beq
  \begin{pmatrix}
    E_{\S^0}  & E_{\S^0\L} \\
    E_{\L \S^0} & E_\L 
    \end{pmatrix},
    \label{eq:LSEmatrix}
\eeq
where the off-diagonal entries are also given by 
Eq.~\eqref{eq:energies}, 
being careful to note that 
$M_B$
is zero for such entries.%
\footnote{
With the mass splittings taken at their physical values in loop diagrams, 
the off-diagonal elements of this matrix are not identical but differ very slightly. 
We set
$E_{\L \S^0} \equiv E_{\S^0 \L}$
in our computation to avoid this complication. 
} 
The eigenstates, 
which we write as 
$\lambda_\pm$, 
are determined from diagonalizing this energy matrix. 
We follow%
~\cite{Parreno:2016fwu}
and define the linear transformation between the eigenstates as
\beq
\begin{pmatrix}
\lambda_+
\\
\lambda_-
\end{pmatrix}
=
\begin{pmatrix}
\phantom{-} \cos \theta & \sin \theta \\
- \sin \theta & \cos \theta
\end{pmatrix}
\begin{pmatrix}
\S^0
\\
\L
\end{pmatrix}
\label{eq:lamsigmix}
,\eeq
where the mixing angle is magnetic field dependent, 
$\theta = \theta(B^2)$.

%%%%%%%%%%%%%%%%%%%%%%%%%
\section{Baryon Energies And Three Scenarios}%
%%%%%%%%%%%%%%%%%%%%%%%%%
\label{s:scenarios}

The remaining parameter required to evaluate the magnetic-field dependence of octet baryon energies is the transition magnetic moment between
the decuplet and octet, 
which has been labeled by 
$\mu_U$
above. 
As is well known in the small-scale expansion, 
see, 
for example, 
Ref.~\cite{Hemmert:1996rw}, 
the largeness of this moment presents a complication in the determination of the magnetic polarizabilities of the nucleon. 
Hence, 
the magnetic-field dependence of baryon energies will inherit a related complication. 
We explore three scenarios for this coupling:
large mixing with decuplet states, 
mitigation by using consistent kinematics, 
and promotion of higher-order counterterms. 
In the first scenario, 
we additionally discuss determination of 
$\mu_U$
using recent experimental results. 
Values of the magnetic polarizability are discussed in the second and third scenarios.

%%%%%%%%%%%%%%%%%%%%%%%
\subsection{Large Decuplet Mixing}             %
%%%%%%%%%%%%%%%%%%%%%%%
\label{s:mixy}

The baryon transition magnetic moment, 
$\mu_U$, 
can be determined using the measured values for the electromagnetic decay widths of the decuplet baryons. 
Beyond the
$\Delta \to N \gamma$
decay, 
recent experimental measurements have been carried out for the electromagnetic widths of the decays
$\Sigma^{*0} \to \L \gamma$%
~\cite{Keller:2011nt}
and
$\Sigma^{*+} \to \Sigma^+ \gamma$%
~\cite{Keller:2011aw}. 
Using the 
magnetic dipole operator appearing in 
Eq.~\eqref{eq:trans}, 
the decay width is found to be
\beq  \label{eq:width}
\Gamma (T \to B \gamma)
=
\a_T
\frac{\o^3 }{2 \pi}  \frac{M_B}{M_T} \left( \frac{e \, \mu_U}{2 M_N} \right)^2
,\eeq
assuming that the electric quadrupole contribution is negligible. 
In the formula for the width: 
$\a_T$
is the relevant
$U$-spin symmetric coefficient appearing in 
Table~\ref{tab:one}; 
the baryon transition moment 
$\mu_U$
appears in nuclear magneton units; 
and, 
$\o$
is the photon energy, 
which is given by 
\beq
\o = \frac{M_T^2 - M_B^2}{2M_T} 
\label{eq:omega}
.\eeq  
The factor of 
$M_B / M_T$
arises from an exact treatment of the relativistic spinor normalization factors. 
Using the three experimentally measured widths, 
we obtain the values
\begin{eqnarray}
\Gamma(\D \to N \g) = 0.660(60) \, \texttt{MeV}
\quad \Rightarrow \quad
 \mu_U &=& 6.04 (27) \, \phantom{1}   \texttt{[NM]}
,\notag \\ 
\Gamma(\S^{* 0} \to \L \g) = 0.445(102) \, \texttt{MeV}
\quad \Rightarrow \quad
\mu_U &=& 6.10 (70) \, \phantom{1} \texttt{[NM]}
,\notag \\
\Gamma(\S^{* +} \to \S^+ \g)  = 0.250(70) \, \texttt{MeV}
\quad \Rightarrow \quad
\mu_U &=& 6.09 (109) \,  \texttt{[NM]}
\label{eq:muTs}
.\end{eqnarray}
Carrying out a weighted fit, 
we obtain the central value
$\mu_U=6.05 \, \texttt{[NM]}$. 
As our analysis is not precise enough to make definite conclusions, 
we will not propagate the uncertainty on this or other parameters. 
The values obtained for 
$\mu_U$
are completely consistent with 
$U$-spin symmetry, 
further consequences of which have been explored in 
Ref.~\cite{Keller:2013hza};
values are also consistent with the na\"ive constituent quark model.

Accounting for the normalization convention used in 
Eq.~\eqref{eq:trans}, 
the transition moment obtained in nuclear magneton units is rather large. 
Assuming there are no corrections that mitigate the size of this coupling, 
we assess whether decuplet-octet mixing in magnetic fields may need to be treated non-perturbatively. 
To perform this assessment, 
we focus on the magnetic moment operators in each coupled system of 
$I_3 \neq 0$
baryons, 
whose members we label by 
$T$ 
and
$B$.%
\footnote{
The 
$I_3 = 0$
octet baryons, 
$\S^0$ 
and
$\L$,
both mix with     
$\S^{*0}$, 
leading to a coupled three-state system,  
which is detailed in 
Appendix~\ref{s:B}. 
}
As 
$U$-spin symmetry forbids the 
$\S^{* -}$--$\S^{-}$
and
$\Xi^{* -}$--$\Xi^{-}$
baryon transitions, 
we omit these baryons from our consideration. 
Their transition moments, 
which are 
\emph{not} 
proportional to 
$\mu_U$,
are expected to be quite small.

For a spin-half baryon 
$B$, 
the dipole transition operator in 
Eq.~\eqref{eq:trans}
leads to mixing between 
$T$ 
and
$B$
baryons in magnetic fields, 
where the 
$I_3$
quantum number of 
the spin three-half
$T$ 
baryon is the same as
$B$. 
Only the 
$m = \pm \frac{1}{2}$
spin states of the 
$T$, 
furthermore, 
can mix with the corresponding 
spin states of the 
$B$. 
Considering the magnetic moment operators in this system, 
the Hamiltonian takes the form
\beq
H
=
\begin{pmatrix}
M_T + \frac{|Q e B|}{2 M_T} & 0 \\
0 & M_B + \frac{|Q e B|}{2 M_B}
\end{pmatrix}
- 
\frac{e B}{2 M_N}
\begin{pmatrix}
2 m \, \mu_{T} &  \mu_{TB} \\
\mu_{TB} &  2 m \, \mu_{B}
\end{pmatrix}
\label{eq:Hmix}
,\eeq 
in the basis
$\begin{pmatrix} T \\ B \end{pmatrix}$, 
where
$m$
denotes the baryon spin state.  
Baryons are assumed to be in their lowest Landau levels, 
where appropriate. 
We have additionally written the transition moment as
$\mu_{T B}$,  
which is related to the $U$-spin symmetric moment through the relation
$\mu_{T B}
= 
\sqrt{\a_T} \, \mu_U$, 
for which the sign can be absorbed into the definition of the mixing angle and is hence irrelevant to the energy eigenvalues. 
Notice that all moments are written in terms of nuclear magneton units. 
The magnetic moments of decuplet baryons 
are defined to be coefficients, 
$\mu_{T}$, 
of the interaction term
$- \frac{e}{M_N} \bm{J} \cdot \bm{B}$, 
where
$\bm{J}$
is the spin operator for the decuplet state $T$.

%%%%%%%%%%%%%%%%%%%%%%%%%%%%%%%%%%%%%%%
\begin{figure}
\centering
%%%%%%%%%%%%%%%%%%%%%%%%%%%%%%%%%%%%%%%
\resizebox{\linewidth}{!}{
        \includegraphics[width=6cm]{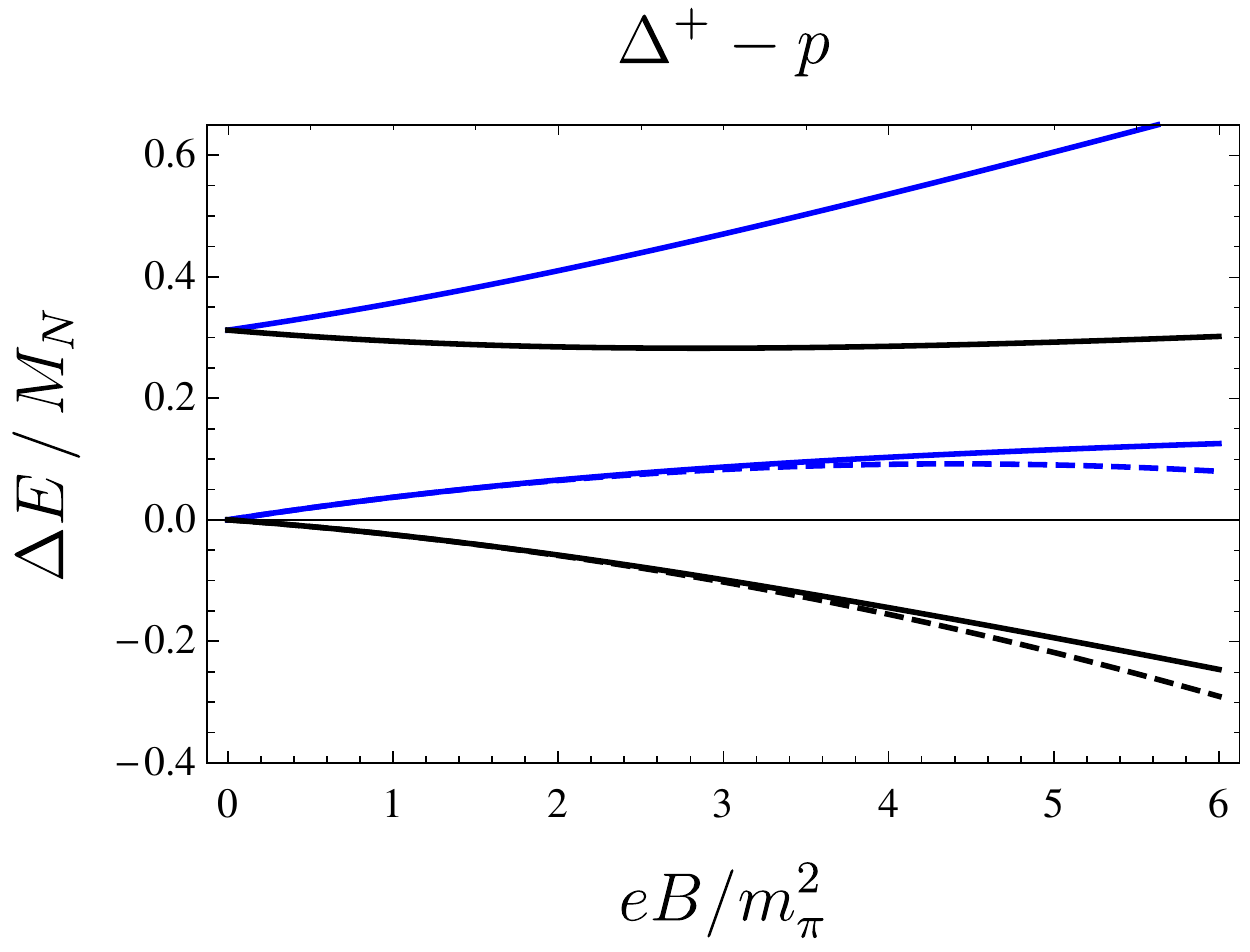}
        $\quad$
        \includegraphics[width=6cm]{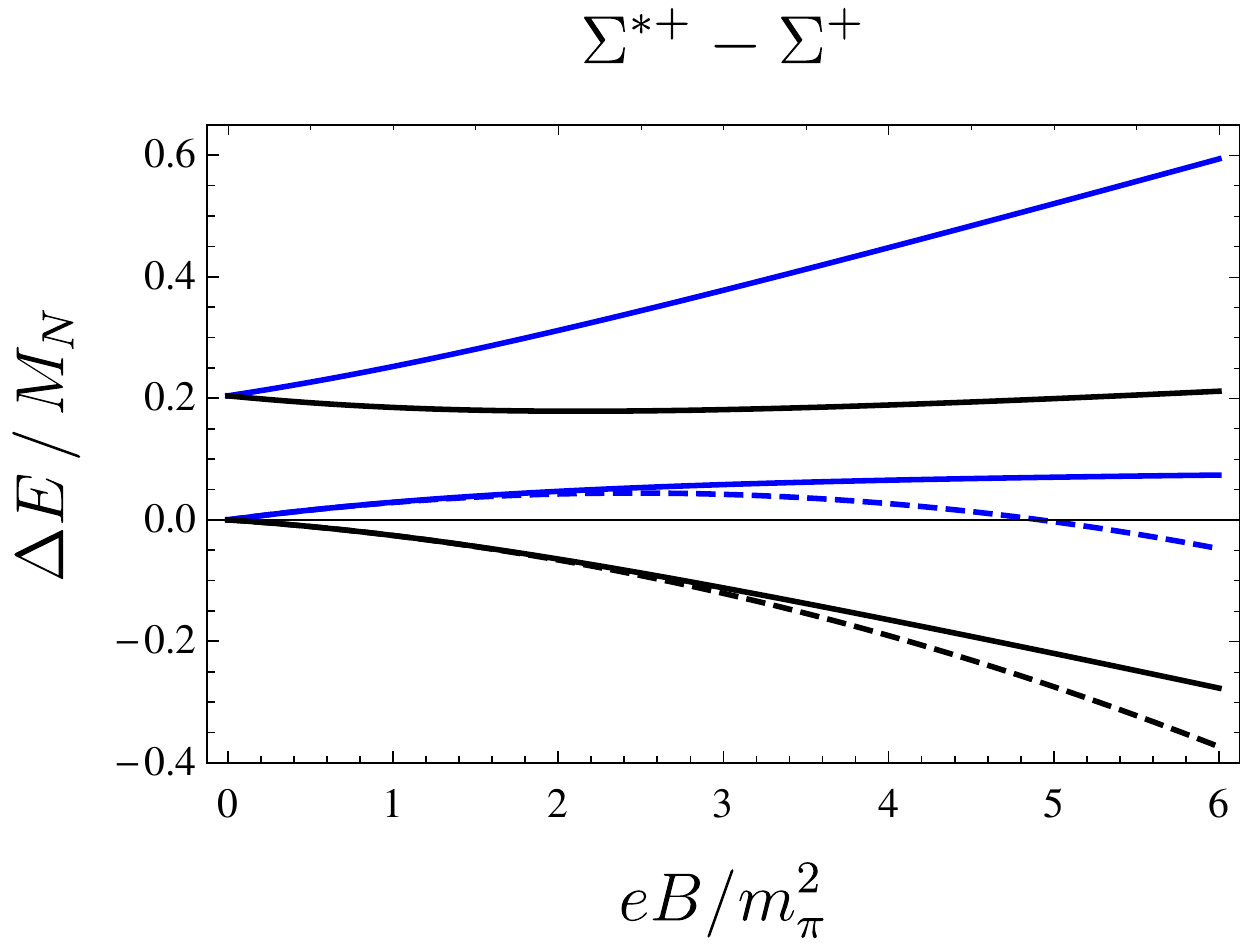}
        }
        \\
        \medskip       
%%%%%%%%%%%%%%%%%%%%%%%%%%%%%%%%%%%%%%%
        \resizebox{\linewidth}{!}{
        \includegraphics[width=6cm]{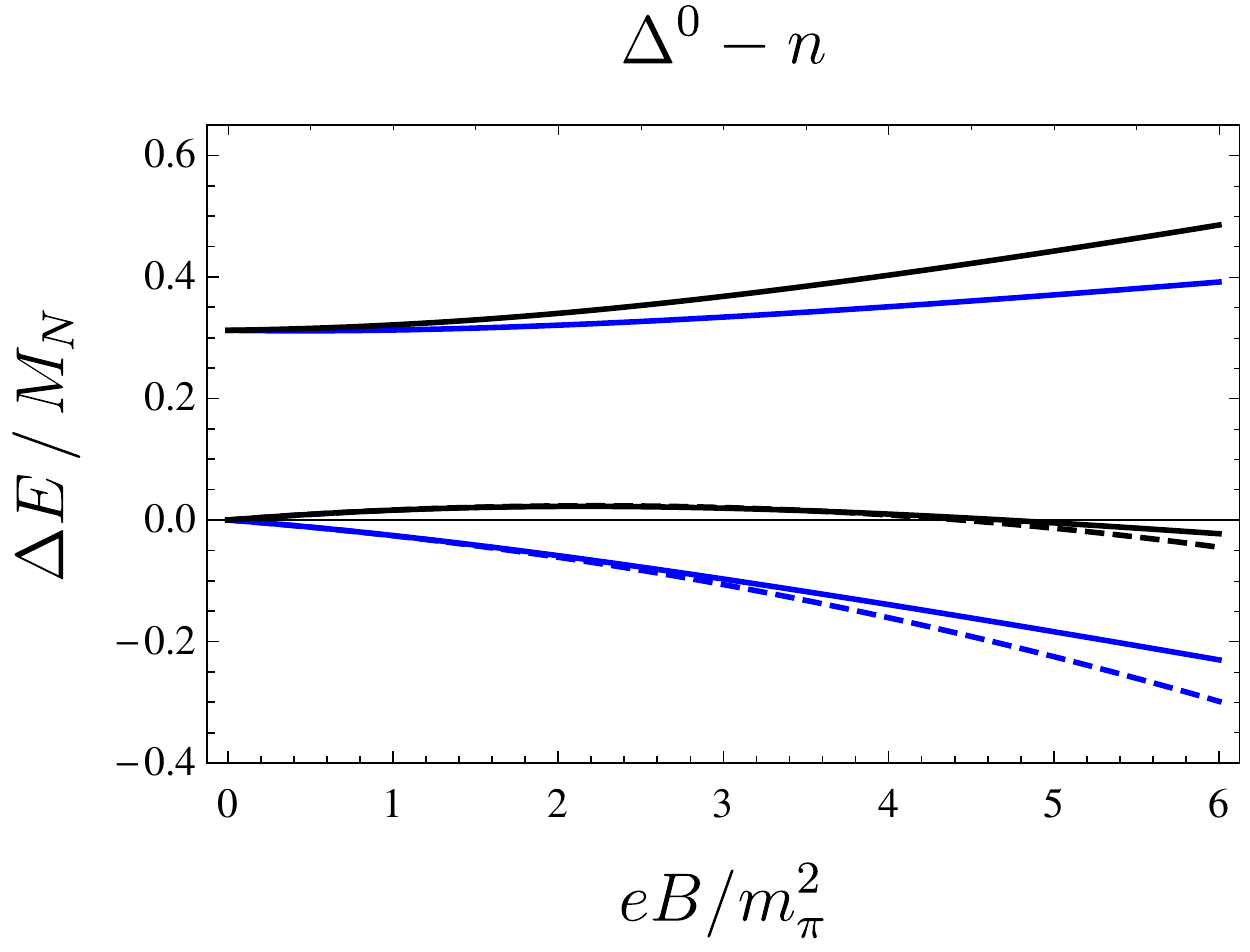}
        $\quad$
        \includegraphics[width=6cm]{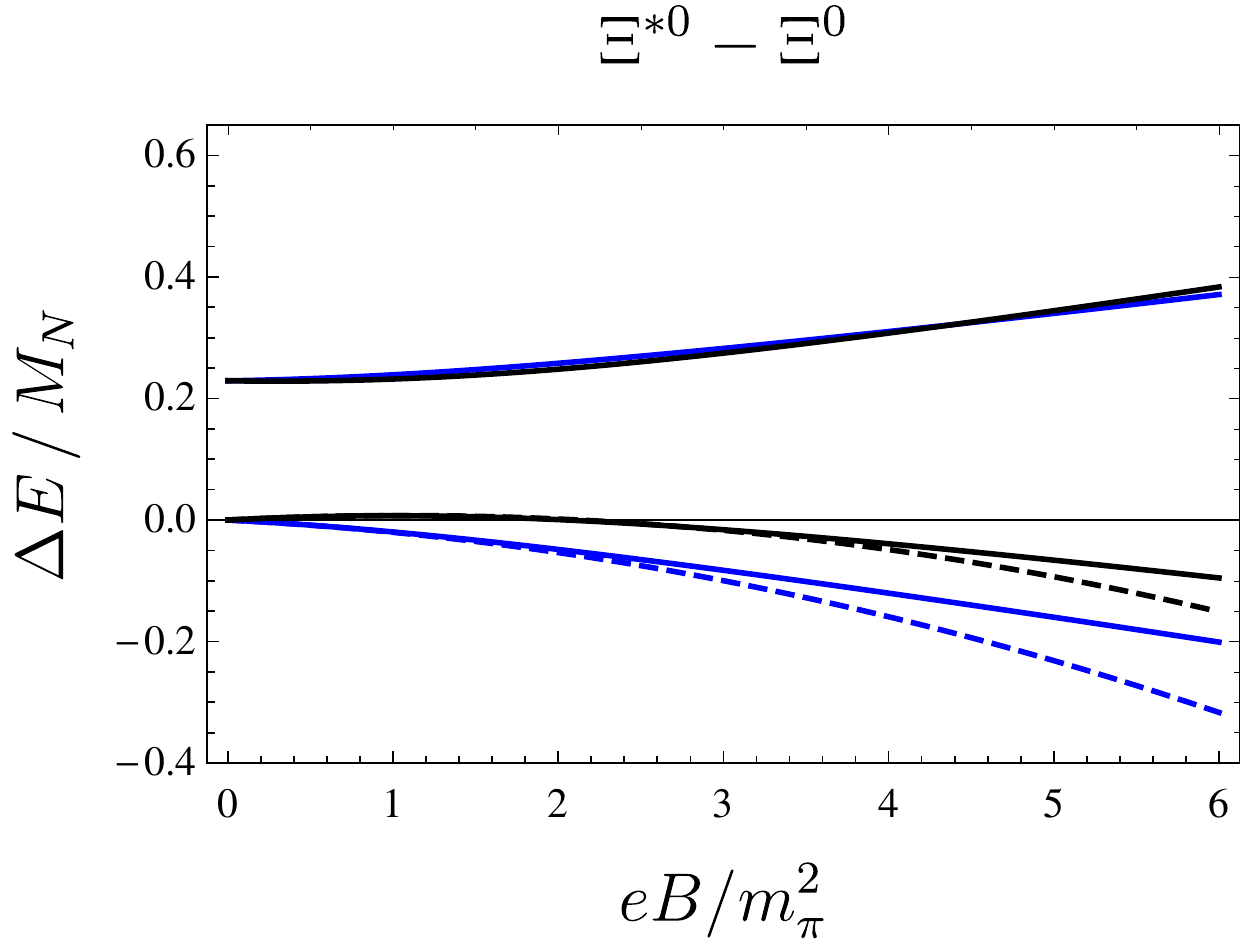}
        }
%%%%%%%%%%%%%%%%%%%%%%%%%%%%%%%%%%%%%%%
\caption{
Assessment of decuplet-octet mixing arising from the Hamiltonian in 
Eq.~\eqref{eq:Hmix}. 
The spin-up (black) and spin-down (blue) eigenstate energies are plotted as a function of the magnetic field. 
These energy shifts, 
$\D E =  E - M_B$,  
are given in units of the nucleon mass, 
with the solid curves corresponding to the full solution in 
Eq.~\eqref{eq:NDeltaE}, 
and the dashed curves representing the approximation in 
Eq.~\eqref{eq:tree}, 
which retains mixing only perturbatively through the decuplet-pole contribution. 
}
        \label{fig:ndmix}
\end{figure}
%%%%%%%%%%%%%%%%%%%%%%%%%%%%%%%%%%%%%%%

From the Hamiltonian in 
Eq.~\eqref{eq:Hmix}, 
the energy eigenvalues for spin states 
$m = \pm \frac{1}{2}$
are
\beq
E^{(m)}_\pm
=
M_B
+ 
\frac{|Q e B|}{2 M_B}
+
\frac{1}{2} 
\left[
E_\D
- 
\mu_+
\frac{e B \, m}{M_N}
\pm
\sqrt{
\left( E_\D -   \mu_-  \frac{e B \, m}{M_N} \right)^2 
+ 
\left(  \mu_{T B}  \frac{e B}{M_N} \right)^2
}
\, \right]
\label{eq:NDeltaE}
,\eeq
where the spin-independent energy difference 
$E_\D$
is given by 
\beq
E_\D = M_T + \frac{|QeB|}{2 M_T} - M_B - \frac{|QeB|}{2 M_B}
,\eeq
and
the parameters 
$\mu_\pm$
are sums and differences of the baryon magnetic moments,  
namely
\beq
\mu_\pm = \mu_T \pm \mu_B
.\eeq 
In the weak-field limit, 
the two spin states of lower energies, 
$E^{(m)}_-$, 
reduce to those determined in 
Eq.~\eqref{eq:tree} 
for the octet baryon $B$, 
with the magnetic moment replaced by its physical value and the 
$\c O(B^2)$
contribution identical to that from the corresponding
decuplet-pole diagram. 
For large 
$\mu_U$
couplings, 
this contribution dominates the magnetic polarizability of the octet state
(see Table~\ref{tab:four} below), 
which is assumed to be the case here necessitating its resummation.

%%%%%%%%%%%%%%%%%%%%%%%%%%%%%%%%%%%%%%%
\begin{figure}
\centering
%%%%%%%%%%%%%%%%%%%%%%%%%%%%%%%%%%%%%%%
\resizebox{\linewidth}{!}{
\includegraphics[height=6cm]{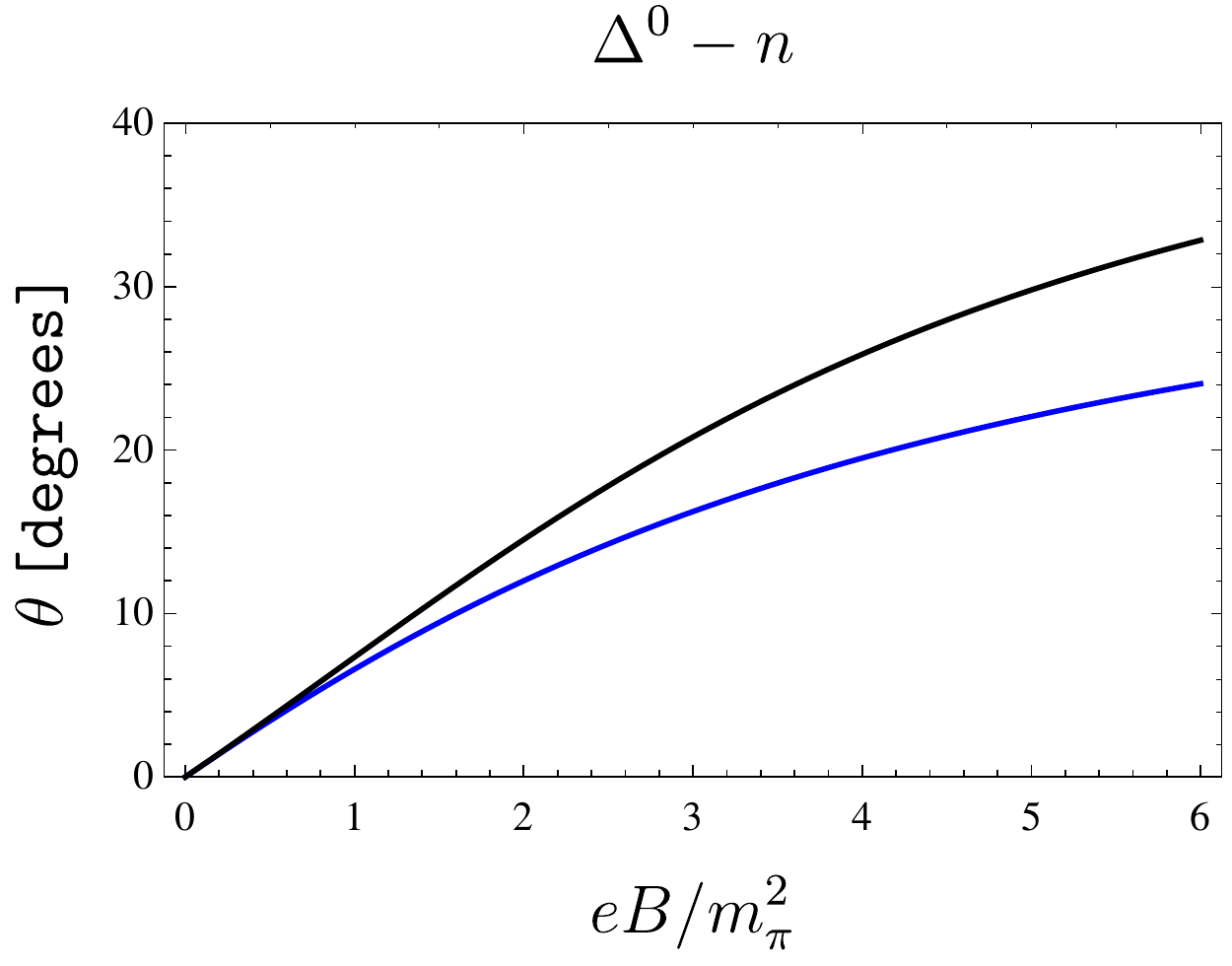}
$\quad$
\includegraphics[height=6cm]{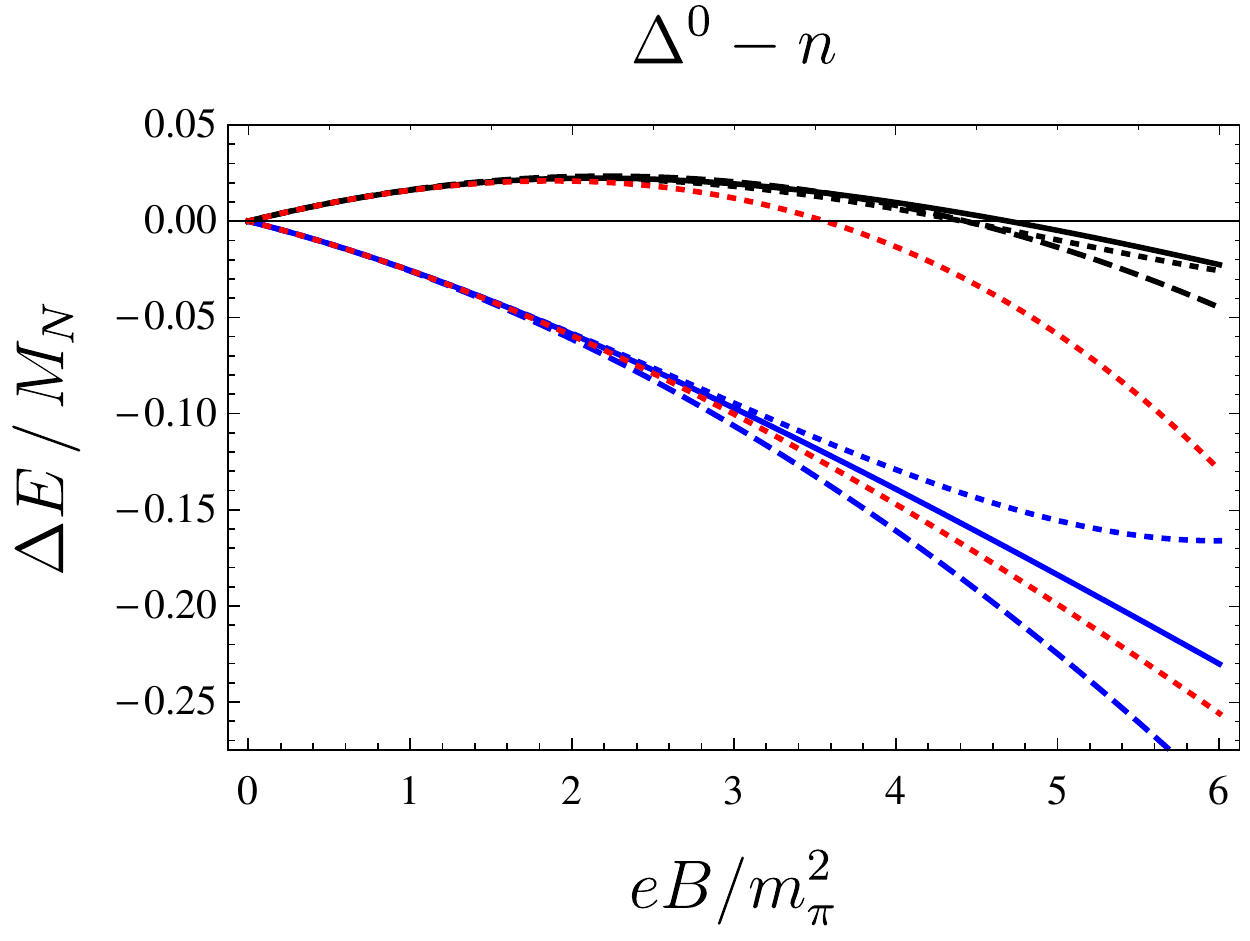}
}
%%%%%%%%%%%%%%%%%%%%%%%%%%%%%%%%%%%%%%%
\caption{
Assessment of perturbative versus non-perturbative mixing in the 
$\D^0$--$n$
system. 
On the left, 
the mixing angle in 
Eq.~\eqref{eq:NDmix}
is plotted for the spin-up (black) and spin-down (blue) states. 
On the right, 
the energy shifts are plotted as a function of the magnetic field, 
with solid curves corresponding to the result in 
Eq.~\eqref{eq:NDeltaE}
for the two spin states. 
Color correlated dashed and dotted curves show the 
$\c O (B^2)$
and
$\c O (B^4)$
approximations to 
Eq.~\eqref{eq:NDeltaE}, 
respectively. 
Despite larger mixing, 
the spin-up energy is well described by perturbation theory in the magnetic field. 
The expansion for the spin-down energy can be considerably improved by 
the resummation into an expansion in  
$\xi$, 
see Eq.~\eqref{eq:Pade}, 
which is shown to 
$\c O (\xi^2)$
for both spin states as the (red) dotted curves. 
In the spin-up case, 
this resummation introduces a pole, 
which is responsible for the rapid divergence from the solid black curve. 
}
\label{f:pert}
\end{figure}
%%%%%%%%%%%%%%%%%%%%%%%%%%%%%%%%%%%%%%%

To evaluate the eigenstate energies, 
values of the decuplet magnetic moments are required. 
A compilation of model and theory results for decuplet moments is contained in the covariant baryon 
$\chi$PT calculation of 
Ref.~\cite{Geng:2009ys}, 
and we adopt the results determined in that particular work:
$\mu_{\D^+} = 2.84 \, \texttt{[NM]}$, 
$\mu_{\S^{*+}} = 3.07 \, \texttt{[NM]}$, 
and
$\mu_{\D^0}  = - \mu_{\Xi^{* 0}} = - 0.36 \, \texttt{[NM]}$, 
which are quite similar to values obtained in the constituent quark model and from large-$N_c$ analyses.
The magnetic field dependence of the energy shifts, 
$\Delta E = E - M_B$,
is shown in 
Fig.~\ref{fig:ndmix}, 
for the 
$I_3 \neq 0$
baryon systems with 
$Q \neq -1$. 
This behavior is compared with that predicted by 
Eq.~\eqref{eq:tree}, 
with good agreement in small fields, 
but with corrections beyond the decuplet-pole contribution required with increasing magnetic field.

While the decuplet pole seems to be a reasonable approximation in most cases, 
we explore, 
in Fig.~\ref{f:pert},
whether mixing with decuplet baryons can be treated perturbatively. 
The figure shows the 
$E^{(m)}_-$  
energy eigenvalue and mixing angle in the 
$\D^0$--$n$
system.
The mixing angle 
$\theta$
is defined by writing this eigenstate, 
$| \L_- \rangle$,  
as the linear combination
\beq \label{eq:NDmix}
| \Lambda_- \rangle 
= 
\cos \theta \, | n \rangle + \sin \theta \, | \D^0 \rangle
.\eeq
Mixing is seen to increase as a function of the magnetic field, 
with greater mixing for the higher-lying spin state. 
Stated this way, 
the feature is generically true across all the baryon systems depicted in 
Fig.~\ref{fig:ndmix}.  
Also shown in 
Fig.~\ref{f:pert} are the 
$\c O (B^2)$
and
$\c O(B^4)$ 
perturbative approximations to the energies, 
where the former is given by 
Eq.~\eqref{eq:tree}. 
For the higher-lying spin state, 
the expansion appears to be under good perturbative control, 
although one should note that the fourth-order expansion includes 
linear, quadratic, and cubic magnetic field terms in addition to the quartic. 
The lower-lying spin state, 
for which the mixing angle is smaller, 
does not exhibit the same convergence properties. 
For this spin state, 
the expansion can be improved by utilizing%
\footnote{
It should be noted that the 
$\xi$
expansion offers little benefit for the case of
$\D^+$--$p$
and
$\S^{*+}$--$\S^{+}$
baryon systems due to the value of their magnetic moment difference,  
$\mu_- \approx 0$. 
The 
$\Xi^{*0}$--$\Xi^{0}$ 
system, 
however,  
is qualitatively the same as the 
$\D^0$--$n$
system, 
and the 
$\xi$
expansion similarly offers an improved scheme for the spin-down state.  
}

\beq
\xi 
= 
\frac{\mu_{TB} \frac{eB}{M_N}}{E_\D - \mu_- \frac{e B m}{M_N}}
\label{eq:Pade}
,\eeq
as the expansion parameter, 
which is suggested by the exact solution in 
Eq.~\eqref{eq:NDeltaE}. 
In the case of 
$\D^0$--$n$
mixing for the spin-down state, 
the resummed delta-pole contribution at 
$\c O (\xi^2)$
provides a much improved approximation to the exact solution, 
as can be seen in 
Fig.~\ref{f:pert}. 
Due to the sign of the spin, 
we have
$\xi < \mu_{TB} \frac{eB}{\D M_N}$, 
which ensures better convergence over an expansion perturbative in 
$B$. 
For the spin-up state, 
by contrast, 
we expect comparatively poor behavior due to  
$\xi > \mu_{TB} \frac{eB}{\D M_N}$. 
The expansion is actually much worse because 
$\xi$ 
exhibits a pole at the magnetic field strength 
$ e B = 2 \D M_N / \mu_- > 0$, 
which noticeably influences the behavior of the 
$\c O(\xi^2)$
expansion in 
Fig.~\ref{f:pert}.

Using the value of 
$\mu_U$
obtained from electromagnetic decays of the decuplet baryons, 
the magnetic mixing of decuplet and octet baryons may need to be treated beyond perturbation theory. 
Assessing the mixing using the linear-order Hamiltonian in 
Eq.~\eqref{eq:Hmix}, 
the decuplet-pole contribution to the energies appears to be a reasonable approximation for magnetic fields satisfying 
$ e B / m_\pi^2 \lesssim 3$. 
While we will ultimately adopt a value for 
$\mu_U$
smaller than that obtained in this section, 
we will nevertheless estimate the effects of mixing beyond the pole term in our complete analysis
(see Sec.~\ref{s:energies} below).

%%%%%%%%%%%%%%%%%%%%%%%%%%%%%%%%%%%%%
\subsection{Magnetic Polarizabilities and Consistent Kinematics}             %
%%%%%%%%%%%%%%%%%%%%%%%%%%%%%%%%%%%%%
\label{s:pols}

The largeness of 
$\mu_U$
may require that baryon mixing be treated non-perturbatively. 
The magnitude obtained above, 
however, 
is unlikely due to the size of contributions from decuplet pole diagrams to magnetic polarizabilities. 
The magnetic polarizabilities appear as the second-order terms in the expansion of energies as a function of the magnetic field. 
In the standard convention, 
the spin-averaged energy, 
$\ol E$,  
has the behavior
\beq
\ol E 
= 
M  
+ 
\frac{|Q e B|}{2 M} 
- 
\frac{1}{2} 4 \pi \b_M B^2 + \cdots
,\eeq 
where the
$\cdots$
represents higher-order terms in the magnetic field strength, 
and the coefficient
$\b_M$
defines the magnetic polarizability.  
Using the spin-independent energy determined in 
Eq.~\eqref{eq:energies}, 
we obtain the magnetic polarizability of the octet baryons
\beq
\beta_M
=
\beta^\text{lp}
+ 
\beta^\text{tr} 
,\eeq
which has been written as the sum of  loop and  tree-level contributions. 
The loop contribution is determined by expanding 
$\d E_2$
in
Eq.~\eqref{eq:main} 
to second order in the magnetic field, 
which leads to the expression
\beq
\beta^\text{lp}
=
\frac{\alpha}{6}
\sum_\c B
\frac{\c A_\c B \, \c S_{2 \c B}}{m_\phi (4 \pi f_\phi)^2}
\,
\c G \left( \frac{\D_\c B}{m_\phi} \right)
,\eeq
with 
$\a$
as the fine-structure constant, 
and the loop function defined by
\beq
\c G (y)
=
- \frac{1}{\sqrt{y^2 - 1}}
\log \left(
\frac{y - \sqrt{y^2 - 1 + i \e}}{y + \sqrt{y^2 - 1+ i\e}} 
\right)
,\eeq 
which takes the particular value 
$\c G(0) = \pi$. 
The tree-level contribution arises from the decuplet pole diagram, 
and has the form%
\footnote{For the
$I_3 = 0$
baryons, 
$\L$
and
$\S^0$, 
there is an additional tree-level contribution to polarizabilities arising from expanding the eigenstate energies determined from 
Eq.~\eqref{eq:LSEmatrix}
to second order in the magnetic field. 
These additional contributions are given by
\beq
\delta \b_\L^\text{tr}
&=&
\phantom{-}
\frac{1}{2\pi\Delta_{\S \L}}\left(\frac{e \, \mu_{\Sigma^0 \L}}{2M_N}\right)^2
,\notag \\
\delta \b_{\S^0}^\text{tr}
&=&
-
\frac{1}{2\pi\Delta_{\S \L}}\left(\frac{e \, \mu_{\Sigma^0 \L}}{2M_N}\right)^2
,\notag 
\eeq
where 
$\D_{\S \L} = M_\S - M_\L$
is the mass splitting. 
Such contributions have the interpretation of pole terms arising from perturbative 
$\S^0$--$\L$
mixing.
Notice that the additional contribution to the 
$\S^0$
polarizability is diamagnetic because the 
$\L$
intermediate state is at a lower energy. 
We do not include these Born-type contributions in our definition of the magnetic polarizabilities of 
$\L$
and
$\S^0$
baryons. 
Instead, 
such contributions are automatically accounted for in the off-diagonal matrix elements of 
Eq.~\eqref{eq:LSEmatrix}, 
and our definition of the polarizabilities then corresponds to the 
$\c O (B^2)$ contribution to the diagonal matrix elements.
}
\beq
\beta^\text{tr}
=
\frac{\alpha_T}{2\pi\Delta_T}\left(\frac{e \, \mu_U}{2M_N}\right)^2
.\eeq
Expressions obtained for nucleon magnetic polarizabilities agree with those determined in Ref.~\cite{Bernard:1991rq,Butler:1992ci,Hemmert:1996rw}.
Furthermore, the octet contributions to hyperon magnetic polarizabilities agree with those determined in Ref.~\cite{Bernard:1992xi}.
Values of these loop and tree-level contributions are given in 
Table~\ref{tab:four}, 
with the corresponding magnetic polarizabilities found from their sum. 
In the case of the nucleon,
the tree-level contribution alone greatly exceeds the experimental values.%
\footnote{
An additional complication is the constraint on the sum of electric and magnetic polarizabilities, 
$\alpha_E + \beta_M$, 
provided by the Baldin sum rule, 
see, 
for example,
Ref.~\cite{Babusci:1997ij}. 
Given that the values we obtain for nucleon electric polarizabilities are consistent with experiment, 
see 
Table~\ref{tab:three}, 
the large magnetic polarizabilities calculated violate the Baldin sum rules for the proton and neutron. 
}

%%%%%%%%%%%%%%%%%%%%%%%%%%%%%%%%%%%%%%%%
\begin{table}
\def\arraystretch{1.33}
\begin{center}
\caption{Possible anatomy of the octet baryon magnetic polarizabilities. 
The various contributions are:
$\b^\text{lp}$, 
the loop contribution;
$\b^\text{tr}$, 
the tree-level pole diagrams;
$b^\text{tr}$, 
the rescaled decuplet-pole diagrams using consistent kinematics; 
$\b^\text{ct}$, 
the promoted counterterms;
and, 
$b^\text{ct}$,
the counterterms relevant for the rescaled decuple-pole diagrams. 
Each contribution is given in units of 
$10^{-4} \, \texttt{fm}^{3}$, 
and counterterms have been determined using the nucleon magnetic polarizabilities as input. 
For this reason, 
those values are starred. 
Finally, 
the 
$\S^0$--$\L$
transition polarizability is not a true polarizability. 
Strictly speaking, 
it represents an off-diagonal matrix element in the system of 
$I_3 = 0$
baryons at second order in the magnetic field. 
} 
\label{tab:four}
\medskip
\begin{tabular}{ |L|||C||C|C|C|C|||C|C|C|C||C|||  }
\hline
\hline
B
& 
\b^\text{{lp}} 
& 
\b^\text{tr} 
& 
b^\text{tr}
&
\b^\text{ct}
&
b^\text{ct}
& 
\b^\text{lp} + \b^\text{tr} 
&
\b^\text{lp} + b^\text{tr}
&
\b^\text{lp} + \b^\text{tr} + \b^\text{ct}
&
\b^\text{lp} + b^\text{tr} + b^\text{ct}
& 
\b_M^\text{expt.}
\tabularnewline
\hline 
p 
&  
\phantom{-}
1.37 
& 
\phantom{-}
13.22 
& 
\phantom{-1}
6.89
&
-12.09
&
\, \, \,
-5.76
& 
\phantom{-}
14.59
& 
\phantom{-1}
8.26
& 
\phantom{i}
* 
2.50
&
\phantom{i}
*
2.50
&
2.5(4)
\phantom{1}
\tabularnewline
n 
&  
\phantom{-}
1.32 
& 
\phantom{-}
13.22 
&
\phantom{-1}
6.89
&
-10.84
&
\, \, \,
-4.51
&  
\phantom{-}
14.54 
& 
\phantom{-1}
8.21
& 
\phantom{i}
*
3.70
&
\phantom{i}
*
3.70
&
3.7(12)
\tabularnewline
\Lambda  
&
\phantom{-}
0.83
&
\phantom{-}
10.79 
&
\phantom{-1}
6.40
&
-16.26
&
\, \, \,
-6.76
&
\phantom{-}
11.62
&
\phantom{-1}
7.23
&
\, \, \,
-4.63
&
\phantom{-1}
0.47
&
-
\tabularnewline
\Sigma^{+}
& 
\phantom{-}
0.96 
& 
\phantom{-}
20.22 
&
\phantom{-}
14.05
&
-12.09
&
\, \, \,
-5.76
& 
\phantom{-}
21.18
&
\phantom{-}
15.02
&
\phantom{-1}
9.09
&
\phantom{-1}
9.26
& 
-
\tabularnewline
\Sigma^{0}
&
\phantom{-}
0.83 
&
\phantom{-1}
5.06 
&
\phantom{-1}
3.51
&
-27.10
&
-11.27
&
\phantom{-1}
5.89
&
\phantom{-1}
4.35
&
-21.21
&
\, \, \,
-6.93
& 
-
\tabularnewline
\Sigma^{-}
&
\phantom{-}
0.71 
&
\phantom{-1}
0.00
&
\phantom{-1}
0.00
&
\phantom{-1}
-
& 
\phantom{-1}
-
&
\phantom{-1}
0.71 
&
\phantom{-1}
0.71
&
\phantom{-1}
-
&
\phantom{-1}
-
&
-
\tabularnewline
\Xi^{0} 
&
\phantom{-}
0.51 
& 
\phantom{-}
18.01 
& 
\phantom{-}
12.45
&
-10.84
&
\, \, \,
-4.51
&
\phantom{-}
18.52
&
\phantom{-}
12.96
&
\phantom{-1}
7.68
&
\phantom{-1}
8.45
& 
-
\tabularnewline
\Xi^{-}  
& 
\phantom{-}
0.30 
&
\phantom{-1}
0.00
&
\phantom{-1}
0.00
&
\phantom{-1}
-
& 
\phantom{-1}
-
&
\phantom{-1}
0.30
&
\phantom{-1}
0.30
&
\phantom{-1}
-
&
\phantom{-1}
-
&
- 
\tabularnewline
\Sigma^{0}\to\Lambda  
& 
-0.03 
&
\,\,\,
-8.76 
& 
\, \, \, 
-6.09
&
\, \, \, 
-9.30
&
\, \, \,
-3.91
&
\,\,\,
-8.79
&
\, \, \,
-6.12
&
-18.18
&
-10.03
& 
-
\tabularnewline
\hline
\hline
\end{tabular}
\end{center}
\end{table}
%%%%%%%%%%%%%%%%%%%%%%%%%%%%%%%%%%%%%%%%

To mitigate the size of the tree-level contribution, 
we note that the normalization factor 
$M_B / M_T$
appearing in the determination of 
$\mu_U$
from the decay width in 
Eq.~\eqref{eq:width} 
has been appended by hand. 
Its removal from the formula is not only consistent with the heavy-baryon power counting, 
it leads to transition moments that show the expected level of 
$SU(3)_V$
breaking. 
Thus, 
the close agreement of the central values of the 
$\mu_U$
parameters in 
Eq.~\eqref{eq:muTs}
for each decay might be accidental. 
Furthermore, 
the formula for the width employs exact kinematics;
whereas, 
to the order we work, 
the photon energy in 
Eq.~\eqref{eq:omega} 
is approximately given by
$\o \approx \D_T$. 
Ordinarily such distinctions are unimportant, 
being of higher order in the expansion, 
however, 
our goal is to expose the sensitivity to such higher-order terms. 
To this end, 
we investigate treating the kinematics consistently within the power counting. 
This can be accomplished by multiplying 
$\mu_U^2$
by the factor 
$\gamma$
defined by
\beq
\gamma
=
\frac{M_B}{M_T} \left(\frac{\o}{\D_T}\right)^3
= 
\frac{M_B}{M_T}
\left( \frac{M_T + M_B}{2 M_T} \right)^3
.\eeq
The corresponding tree-level contributions to the magnetic polarizability are then given by
$b^\text{tr}=\g \b^\text{tr}$, 
which have been included in Table~\ref{tab:four}.
Notice that consistent kinematics are being employed for the 
$\Delta$--$N$, 
$\Sigma^{*0}$--$\Lambda$, 
and 
$\Sigma^{*+}$--$\Sigma^+$ 
transition moments, 
for which experimental results are available. 
In the case of 
$\Sigma^{*0}$--$\Sigma^0$  
and 
$\Xi^{*0}$--$\Xi^0$  
transitions, 
for which no experimental constraints currently exist, 
we use, as a guess, 
the U-spin symmetry prediction, 
but scaled by $\gamma$ to reduce the size as might be expected from the reaction kinematics.    
While the magnetic polarizabilities of the nucleons are subsequently reduced, 
they still exceed the experimental values.

%%%%%%%%%%%%%%%%%%%%%%%
\subsection{Counterterm Promotion}            %
%%%%%%%%%%%%%%%%%%%%%%%
\label{s:promo}

In the context of the small-scale expansion, 
one solution to the large size of calculated nucleon magnetic polarizabilities is to partially cancel the effect of the delta-resonance pole diagram by promoting 
counterterms from the 
$\c O (\epsilon^4)$
Lagrangian density. 
In two-flavor $\chi$PT, 
there are two such local operators. 
Consequently one can adjust these terms to produce any values for the proton and neutron polarizabilities. 
In the three-flavor chiral expansion, 
the same procedure yields 
$U$-spin relations among the polarizabilities of the baryon octet, 
which are detailed here.

The 
$\c O (\epsilon^4)$
magnetic polarizability operators of the octet baryons are contained in the Lagrangian density
\beq
\Delta\mathcal{L}
=
-\frac{1}{2}4\pi B^2
\sum_{i=1}^4\beta_i^\text{ct} \, \mathcal{O}_i,
\eeq
where the 
$\beta^\text{ct}_i$ 
are numerical coefficients,  
and a basis for the four operators
$\c O_i$
is specified by
\begin{align*}
\mathcal{O}_1&=\Tr \left( \overline{B}B \right) \, \Tr \left(  \c Q^2 \right)
,\\
\mathcal{O}_2&=\Tr \left( \overline{B} \left\{ \c Q, \left\{ \c Q, B \right\} \right\} \right) 
,\\
\mathcal{O}_3&=\Tr \left( \overline{B} \left\{ \c Q, \left[ \c Q, B \right] \right\} \right) 
,\\
\mathcal{O}_4&=\Tr \left( \overline{B} \left[\c Q, \left[\c Q,B \right] \right] \right)
.\numberthis
\end{align*}
These operators were enumerated in 
Ref.~\cite{Parreno:2016fwu} 
in the context of 
$SU(3)_V$ 
symmetric lattice QCD computations;
whereas, 
they enter here as the leading terms in the expansion about the 
$SU(3)_L \times SU(3)_R$
limit. 
Including the 
$\S^0$-$\L$
transition, 
there are nine polarizabilities and only four operators;  
hence, 
there exist five relations between the counterterm contributions to the polarizabilities.  
Three of these are the relations obtained under interchanging the
$d$ 
and 
$s$
quarks, 
namely
\begin{align*}
\beta^\text{ct}_p
=
\beta^\text{ct}_{\Sigma^+},
\qquad
\beta^\text{ct}_n
=
\beta^\text{ct}_{\Xi^0}
,\qquad
\beta^\text{ct}_{\Sigma^-}
=
\beta^\text{ct}_{\Xi^-}
,\numberthis\end{align*}
while the remaining two relations can be chosen as
\beq
\frac{1}{\sqrt{3}}\beta^\text{ct}_{\Sigma^0\Lambda}
&=
\beta^\text{ct}_\Lambda-\beta^\text{ct}_n
=
\frac{1}{2}(\beta^\text{ct}_{\Sigma^0}-\beta^\text{ct}_\Lambda)
.\eeq

In this scenario, 
the experimental values of the nucleon magnetic polarizabilities can thus be employed to determine the counterterm 
contributions to the 
$\Sigma^+$
and
$\Xi^0$
polarizabilities. 
Notice that knowledge of the nucleon magnetic polarizabilities cannot help constrain the counterterms for the 
$\S^-$
and
$\Xi^-$
baryons, 
because  the corresponding 
$\S^{*-}$
and
$\Xi^{*-}$
pole diagrams vanish by 
$U$-spin symmetry. 
In the large-$N_c$ limit, 
the counterterm 
$\b_1^\text{ct}$
vanishes. 
In this limit, 
one obtains an additional relation, 
which can be utilized to show
\beq
\beta^\text{ct}_n
=
\frac{2}{5}\beta^\text{ct}_{\Sigma^0}
=
\frac{2}{3}\beta^\text{ct}_{\Lambda}
=
\frac{2}{\sqrt{3}}\beta^\text{ct}_{\Sigma^0\Lambda}
.\eeq
Thus $U$-spin symmetry along with the large-$N_c$ limit permit us to determine five of the seven octet baryon magnetic polarizabilities,  
using the proton and neutron magnetic polarizabilities for input. 
The nucleon counterterm contribution provides the diamagnetism necessary to cancel large paramagnetic effects from the delta-pole contribution. 
Results for the counterterm contribution and magnetic polarizabilities of octet baryons are given in 
Table~\ref{tab:four}. 
We adopt two possibilities for the decuplet pole contribution:
one uses the baryon transition moment obtained from the full kinematics, 
while the other uses the moment obtained from kinematics expanded consistently in our power counting. 
Results are summarized as follows. 
The magnetic polarizaibilities of the 
$\S^+$
and
$\Xi^0$ 
remain paramagnetic and somewhat large. 
The 
$\Lambda$
polarizability is substantially reduced, 
and is quite small or may even become diamagnetic. 
On the other hand, 
the 
$\Sigma^0$
polarizability becomes considerably diamagnetic in nature. 
The transition polarizability between the 
$\S^0$
and
$\Lambda$
baryons is consistently negative, 
perhaps more so with addition of the counterterms.

%%%%%%%%%%%%%%%%
\subsection{Baryon Energies}%
\label{s:energies}%%%%%%%
%%%%%%%%%%%%%%%

%%%%%%%%%%%%%%%%
%%%%%%%%%%%%%%%%
\begin{figure}[t!]
%%%%%%%%%%%%%%%%
\resizebox{\linewidth}{!}{
\includegraphics[width=6cm]{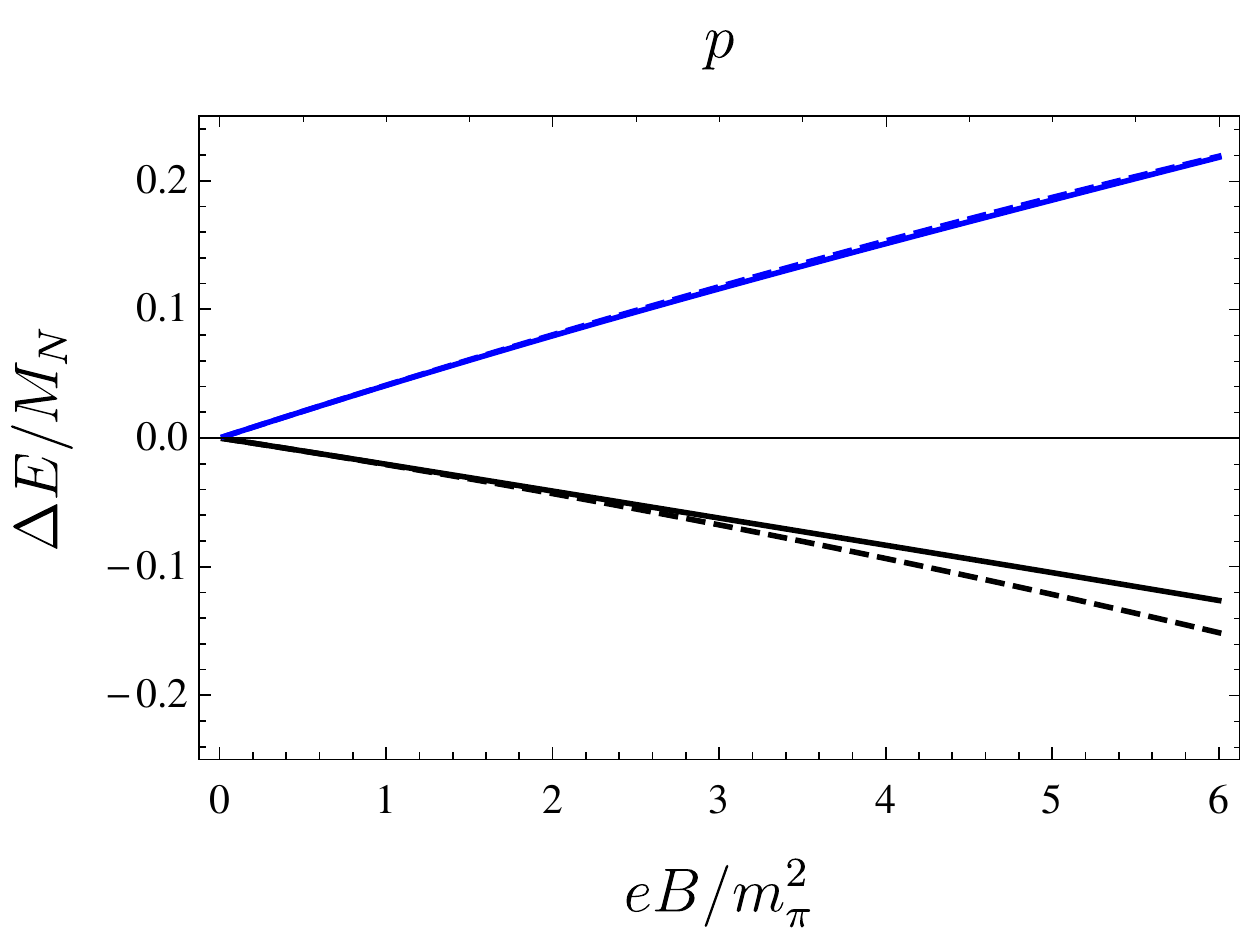}
$\quad$
\includegraphics[width=6cm]{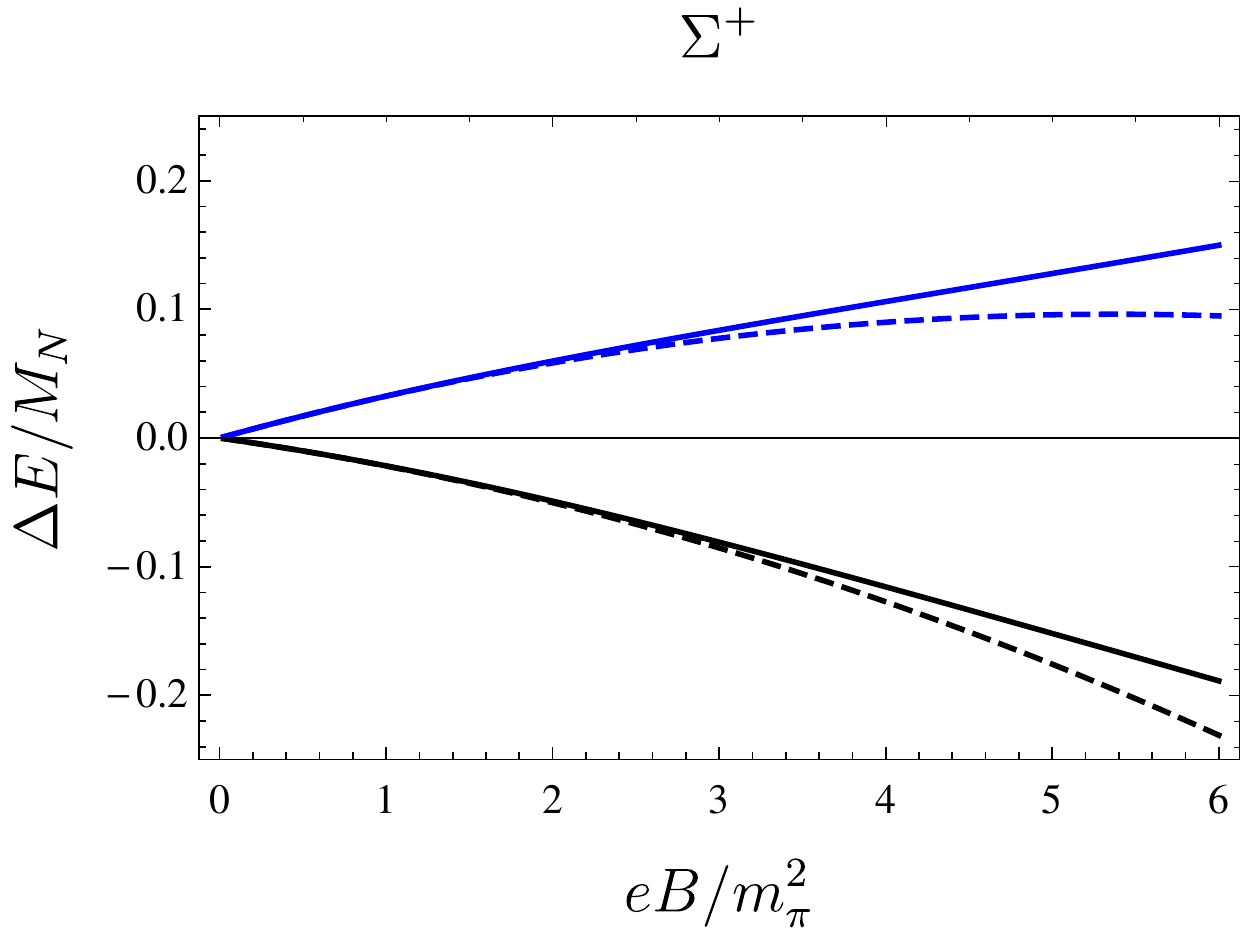}
}
\medskip
\resizebox{\linewidth}{!}{
\includegraphics[width=5cm]{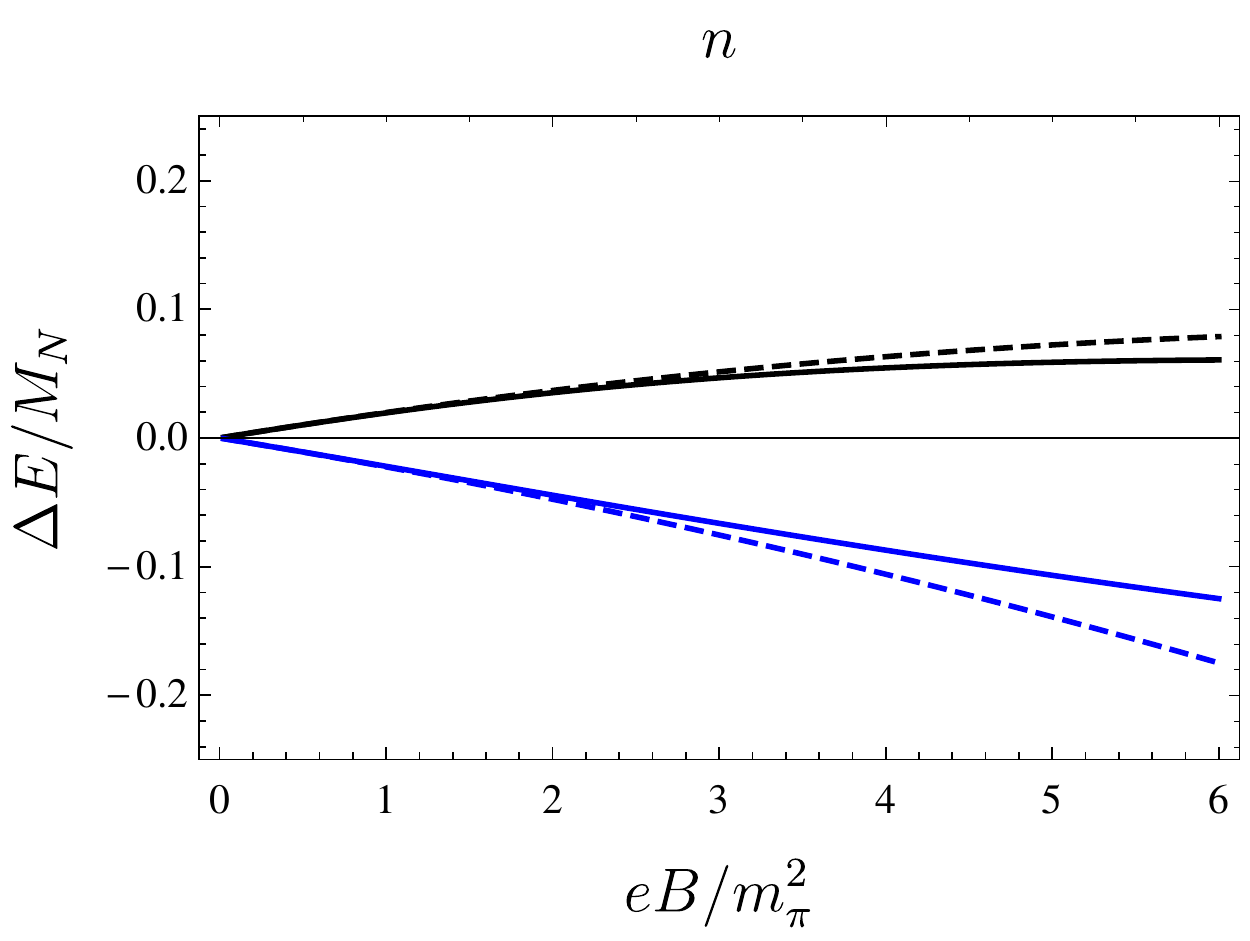}
$\quad$
\includegraphics[width=5cm]{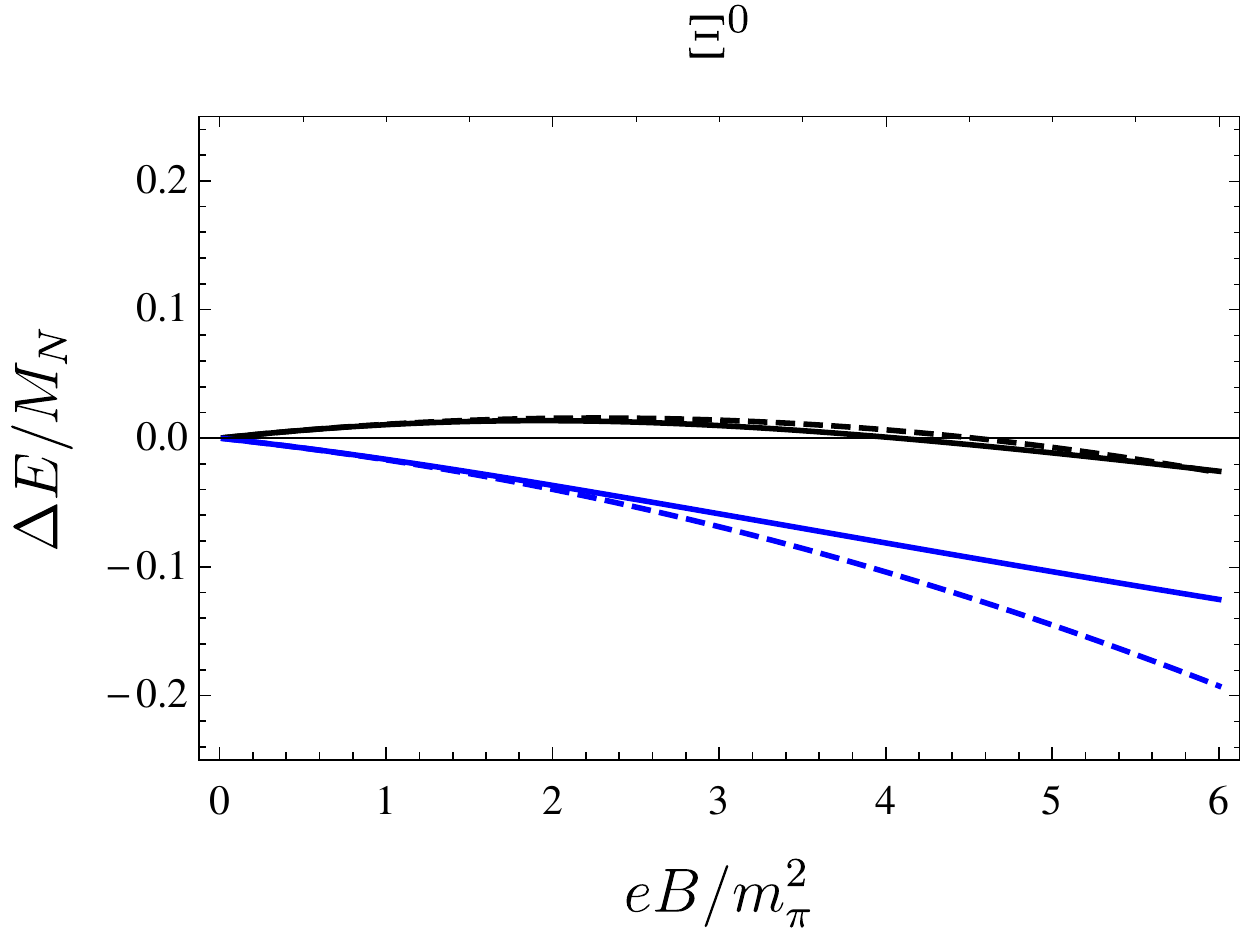}
}
\resizebox{\linewidth}{!}{
\includegraphics[width=5cm]{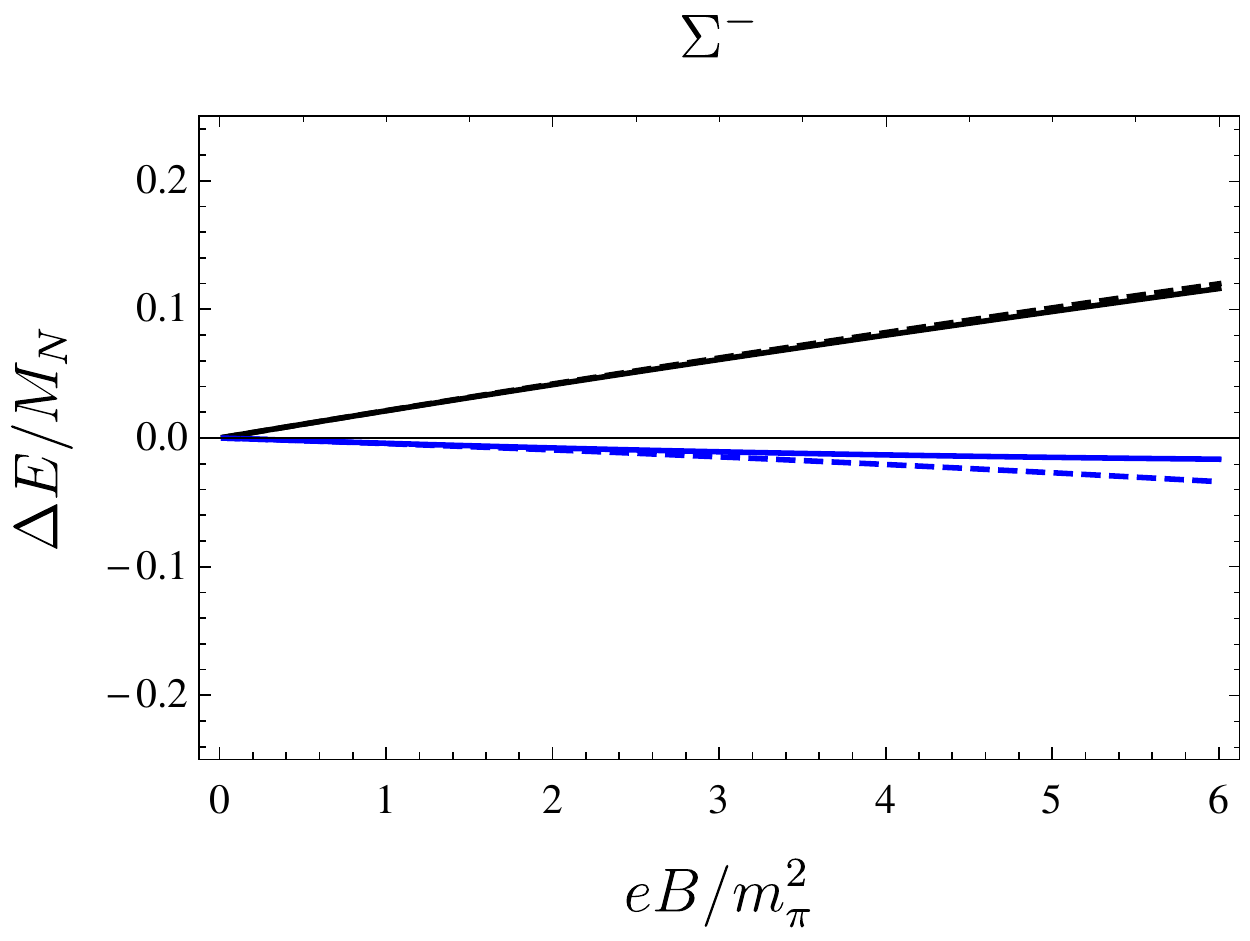}
$\quad$
\includegraphics[width=5cm]{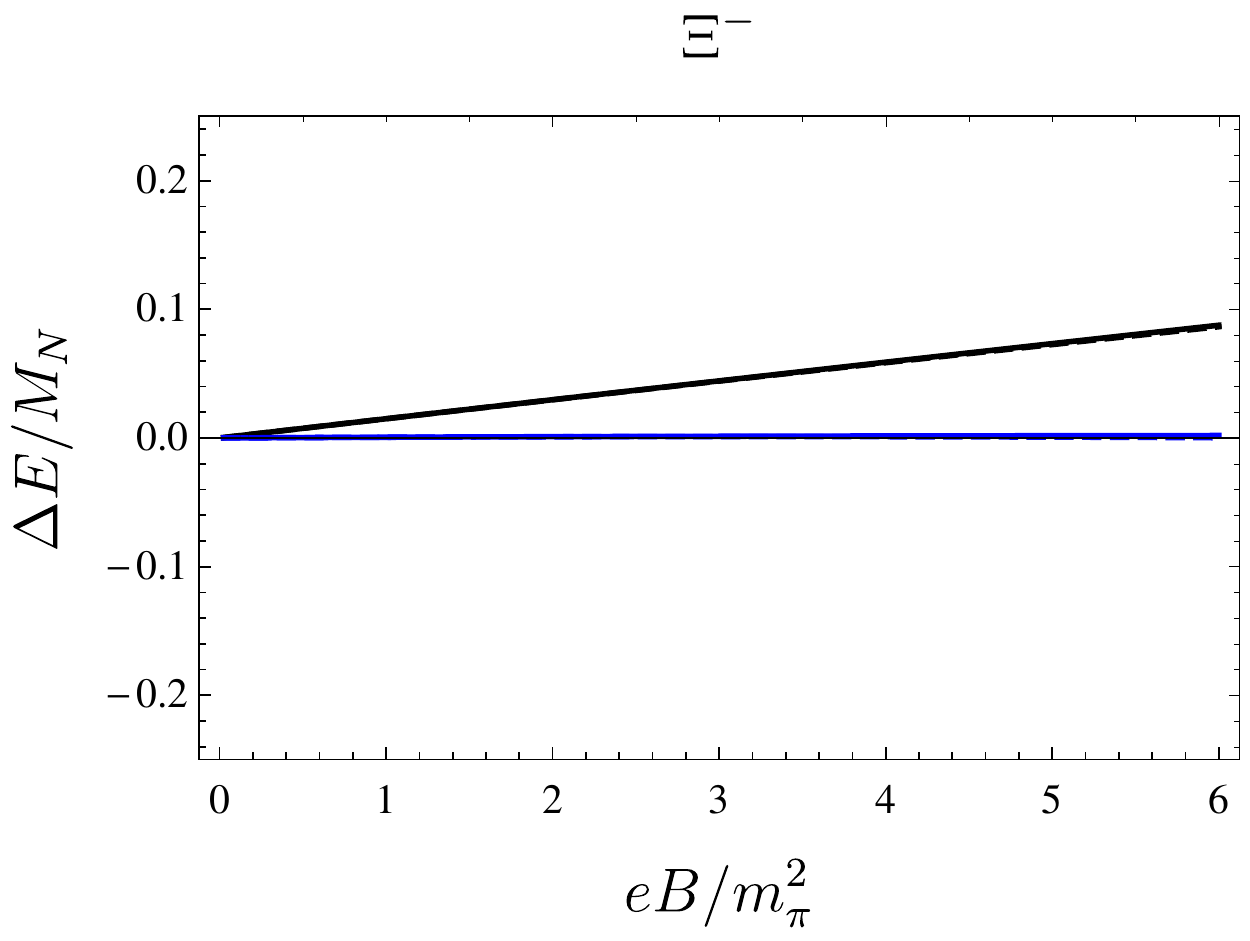}
}
%%%%%%%%%%%%%%%%
\caption{
Magnetic-field dependence of the
$I_3 \neq 0$ 
baryon energy shifts,
$\D E = E - M_B$.  
The black curves correspond to spin-up states, 
while the lighter (blue) curves correspond to spin down. 
The dashed curves only account for effects up to 
$\c O (B^2)$, 
see 
Eq.~\eqref{eq:B2}, 
where the magnetic polarizability is taken to be that in
Eq.~\eqref{eq:betas}.
The solid curves additionally include non-perturbative effects beyond 
$\c O(B^2)$, 
which include loop corrections and mixing with decuplet baryons, 
see 
Eq.~\eqref{eq:H}.   
}
\label{fig:energies}
\end{figure}
%%%%%%%%%%%%%%%%
%%%%%%%%%%%%%%%%

To investigate the magnetic-field dependence of the octet baryon energies, 
we adopt the values of magnetic polarizabilities obtained through counterterm promotion 
using consistent kinematics. 
Thus, 
the baryon magnetic polarizabilities are taken as
\beq \label{eq:betas}
\b_M
\equiv
\b^\text{lp}
+ 
b^\text{tr}
+ 
b^\text{ct}
,\eeq
where the numerical values appear in 
Table~\ref{tab:four}. 
For the proton and neutron, 
these are the experimentally measured ones;
while, 
for the other baryons,
the values are a consequence of 
$U$-spin symmetry and the large-$N_c$ limit. 
Finally, 
the 
$\S^-$
and
$\Xi^-$
magnetic polarizabilities, 
for which counterterm contributions cannot be estimated,  
are taken to be their one-loop values. 
With this choice, 
the octet baryon energy to 
$\c O (B^2)$
is given by 
\beq
\D E
=
\frac{|QeB|}{2 M_B}
- 
m \, \mu_B 
\frac{e B}{M_N} 
- 
\frac{1}{2} 4 \pi \b_M B^2
\label{eq:B2}
,\eeq
where the zero-field result has been subtracted to produce the energy shift,
$\D E = E - M_B$, 
and 
$m = \pm \frac{1}{2}$
for the spin states.

Beyond 
$\c O(B^2)$, 
we have additionally determined loop contributions as non-perturbative functions of 
$|e B| / m_\pi^2$, 
and are able to account for potentially large mixing with decuplet states. 
To incorporate these effects, 
we determine the energy eigenvalues of the Hamiltonian
\beq
H
=
\begin{pmatrix}
\ol E_T & 0 \\
0 & \ol E_B
\end{pmatrix}
- 
\frac{eB}{2M_N}
\begin{pmatrix}
2 m \, \mu_T & \mu_{TB}
\\
\mu_{TB} & 2 m \, \c M_B
\end{pmatrix}
\label{eq:H}
,\eeq 
which goes beyond the approximation used for assessing decuplet mixing in 
Sec.~\ref{s:mixy} 
above. 
The spin-independent entries are defined by 
\beq
\ol E_T 
&=&
M_T + \frac{|QeB|}{2 M_T}
,\notag \\
\ol E_B
&=& 
M_B + \frac{|QeB|}{2 M_B} - \frac{1}{2} 4 \pi b^\text{ct} B^2 + \d E_2
\label{eq:Ebar}
,\eeq
whereas the spin-dependent term
$\c M_B$
is given by
\beq
\c M_B
= 
\mu_B 
+ 
2 M_N \,  \d E_1
\label{eq:MBs}
.\eeq
Notice that the spin-dependent loop contribution
$\d E_1$, 
which is given in
Eq.~\eqref{eq:main},
vanishes in zero magnetic field, 
so that 
$\mu_B$
are the physical baryon magnetic moments in nuclear magneton units. 
The magnetic polarizability appearing in 
$\ol E_B$
requires a subtraction due to the treatment of decuplet mixing.
The decuplet-pole contribution, 
as well as higher-order effects, 
are already generated by the off-diagonal terms in 
Eq.~\eqref{eq:H}, 
which are proportional to the rescaled transition moment, 
$\mu_{TB} = \sqrt{\g \, \a_T} \, \mu_U$.
Thus, use of the rescaled polarizability counterterm
$b^\text{{ct}}$
in 
Eq.~\eqref{eq:Ebar}
ensures that the magnetic polarizabilities are given by 
Eq.~\eqref{eq:betas}.

For each 
$I_3 \neq 0$
baryon, 
we show the magnetic-field dependence of their energies in 
Fig.~\ref{fig:energies}. 
Energy shifts, 
$\D E = E - M_B$,  
are plotted for the spin-up and spin-down states, 
in units of the nucleon mass, 
$\D E / M_N$, 
and grouped according to the baryon's charge. 
For each state, 
moreover, 
we compare the energy computed to 
$\c O (B^2)$,  
given in 
Eq.~\eqref{eq:B2}, 
with the energy including higher-order effects, 
which corresponds to the lower eigenvalue of 
Eq.~\eqref{eq:H}. 
Such higher-order effects arise from charged-meson loops as well as mixing with decuplet baryons, 
and make contributions at 
$\c O (B^3)$
and higher. 
These contributions do not necessarily have a well-behaved expansion in powers of the magnetic field over the range of fields plotted
(see 
Fig.~\ref{fig:dif7},
for the meson loop contributions in particular).  
The loop contributions, 
furthermore, 
are generally largest for the nucleons, 
smaller for the 
$\Sigma$'s, 
and smallest for the 
$\Xi$'s. 
This pattern is to be expected: 
the pion loop contributions dominate over those of the kaon
($\propto f_\pi^{-2} m_\pi^{-1}$ versus $\propto f_K^{-2} m_K^{-1}$ for $\D = 0$), 
and pions couple more strongly to multiplets with lower strangeness. 
In particular, 
the meson loop contributions to the 
$\Xi^0$
and
$\Xi^-$
energies are numerically very small. 
Notice that all plots terminate at the value
$e B / m_\pi^2 = 6$. 
Beyond this value,  
neglected higher-order corrections, 
which na\"ively scale as 
$e B / M_N^2$, 
may be appreciable.

The qualitative behavior of the energies as a function of the magnetic field shown in 
Fig.~\ref{fig:energies}
is somewhat similar when grouped by baryon charge.  
When all effects are accounted for, 
the proton and 
$\S^+$
spin states have a curious linear-appearing behavior. 
Within the 
$\c O (B^2)$
approximation, 
the 
$\Sigma^+$ 
states exhibit some curvature due to the large paramagnetic value assumed for its magnetic polarizability.
The neutron and 
$\Xi^0$
compare similarly:
the neutron spin states exhibit some curvature, 
but not as great as that seen for the 
$\Xi^0$. 
The spin-up neutron state is very insensitive to higher-order corrections due to a near cancellation between loop contributions and $\D^0$ mixing. 
For the 
$\Xi^0$, 
however,
deviations from the 
$\c O(B^2)$
approximation are due almost entirely to mixing with 
$\Xi^{*0}$, 
because the pion-loop contributions are very small. 
The 
$\Sigma^-$
and
$\Xi^-$
exhibit very linear behavior due to the small values assumed for their magnetic polarizabilities and small loop contributions. 
Because the magnetic moments of these baryons are very close to their Dirac values,
see 
Ref.~\cite{Parreno:2016fwu}, 
the spin-down states show an almost exact cancellation of the energy from the lowest Landau level. 
This leads to the near zero shifts for spin-down states observed in the figure.

%%%%%%%%%%%%%%%%%%%%%%%%%%%%%%%%%%%%%%%
\begin{figure}
\centering
\resizebox{0.5\linewidth}{!}{
        \includegraphics[width=6cm]{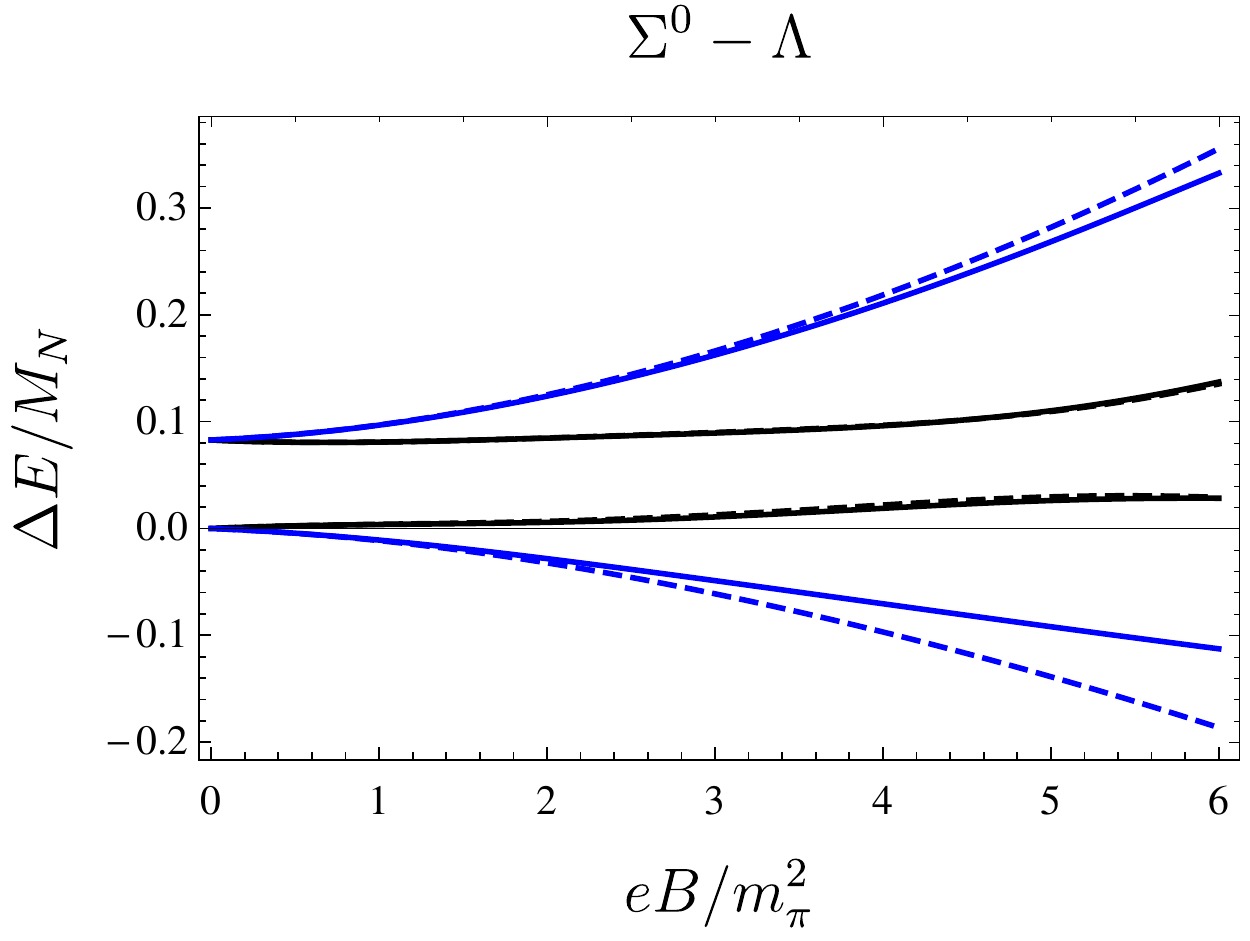}
        }
\caption{Energy eigenvalues in the  
$\S^0$-$\L$ 
system as a function of the magnetic field
for spin-up (black) 
and spin-down (blue) 
states. 
These energies are shifted relative to the mass of the 
$\L$
baryon, 
$\D E = E - M_\L$, 
and plotted in units of the nucleon mass.     
The dashed curves indicate results using the 
$\c O(B^2)$
approximation in
Eq.~\eqref{eq:2state}, 
while the solid curves include contributions from charged-meson loops and mixing with the 
$\S^{*0}$
baryon, 
determined from the Hamiltonian in 
Eq.~\eqref{eq:3state}.   
}
        \label{fig:dif9}
\end{figure}
%%%%%%%%%%%%%%%%%%%%%%%%%%%%%%%%%%%%%%%

Finally we determine the eigenstate energies in the coupled
$\Sigma^0$--$\Lambda$
system by diagonalizing their energy matrix. 
Potentially large mixing with the 
$\S^{*0}$
baryon, 
however, 
renders
the two-state description of 
Eq.~\eqref{eq:LSEmatrix}
insufficient. 
An assessment of mixing with the
$\S^{*0}$
is carried out in 
Appendix~\ref{s:B}, 
and shows that mixing with 
$\S^{*0}$
is nearly as important as 
$\S^0$--$\L$ 
mixing itself. 
The eigenstate energies determined from a coupled three-state analysis are plotted as a function of the magnetic field in 
Fig.~\ref{fig:dif9}. 
The splitting between spin-down states increases as a function of the magnetic field. 
The lower, spin-down eigenstate exhibits a greater dependence on higher-order corrections, 
which arise mainly from mixing with 
$\S^{*0}$. 
On the other hand, 
spin-up eigenstates appear quite insensitive to higher-order effects.

%%%%%%%%%%%%%%%%%%%%%%%%%%%%%%%%%%%%%%%
\section{Summary}
\label{s:summy}
%%%%%%%%%%%%%%%%%%%%%%%%%%%%%%%%%%%%%%%

We determine energies of the octet baryons in large, uniform magnetic fields using heavy baryon $\chi$PT. 
The calculation employs a modified power-counting scheme that treats the magnetic field non-perturbatively compared with the square of the meson mass, 
Eq.~\eqref{eq:1}.
The analytic expressions obtained for baryon energies are summarized in 
Sec.~\ref{s:third}.
These results correspond to 
$\c O(\epsilon^3)$
in the combined heavy baryon and chiral expansion, 
although we have not adhered to a strict power counting. 
Instead, 
we resum a subset of 
$SU(3)_V$
breaking effects. 
For reference, 
octet baryon magnetic moments and electric polarizabilities determined in this scheme are provided in 
Appendix~\ref{s:magpol}.

While evaluation of the magnetic-field dependence of  baryon energies is possible using phenomenological values for the various couplings, 
the transition dipole moment between the decuplet and octet baryons presents a critical issue. 
The value determined from the electromagnetic decays of the decuplet is large enough to require careful treatment of mixing with 
decuplet states in uniform magnetic fields, 
see 
Sec.~\ref{s:mixy}. 
For the $I_3 = 0$ baryons, 
$\S^{*0}$, 
$\S^0$, 
and
$\L$, 
their coupled three-state system is detailed in 
Appendix~\ref{s:B}. 
Adhering to a more strict power counting, 
the extracted value of the baryon transition moment can be reduced, 
as detailed in 
Sec.~\ref{s:pols}. 
This reduction in value, 
however, 
is not enough to produce nucleon magnetic polarizabilities close to their smaller experimental values.
Within the heavy-baryon approach, 
the commonplace solution is to promote higher-order counterterms in order to provide the necessary diamagnetic contributions. 
We detail consequences of this hypothesis for the baryon octet in 
Sec.~\ref{s:promo}, 
using $U$-spin and large-$N_c$ arguments. 
Possible anatomy of the magnetic polarizabilities of the octet baryons is presented in 
Table~\ref{tab:four}. 
The large variation of results over the different scenarios considered does not currently allow for predictions to be made. 
Forcing the nucleon polarizabilities to take on their experimental values, 
however, 
does not rule out large paramagnetic polarizabilities for other members of the octet
(the $\S^+$ and $\Xi^0$ baryons in particular). 
In 
Sec.~\ref{s:energies}, 
we adopt the experimental nucleon magnetic polarizabilities, 
and best guesses for the remaining members of the octet, 
to investigate the magnetic-field dependence of baryon energies, 
see Figs.~\ref{fig:energies} and \ref{fig:dif9}. 
The 
$p$ 
and 
$\S^+$
energies appear remarkably linear after accounting for effects beyond 
$\c O (B^2)$. 
The electrically neutral baryons exhibit a good degree of cancelation of these higher-order effects in the energies of spin-up states, 
but not for their spin-down states. 
The 
$\S^-$
and
$\X^-$
appear rather point-like and rigid,
magnetically speaking.

The results obtained here can be utilized to address the pion-mass and magnetic-field dependence of baryon energies, 
which are relevant for lattice QCD computations of magnetic polarizabilities.
The study of octet baryons in large magnetic fields, 
furthermore, 
provides a diagnostic on the potentially large paramagnetic contributions from decuplet states.  
To this end, 
refined 
$\chi$PT
computations 
(incorporating loop contributions to the baryon transitions, and exploring alternative power-counting schemes for inclusion of the decuplet)
appear necessary. 
In lieu of experimental results for these baryons, 
moreover,
lattice QCD can provide the necessary information to disentangle 
long-range (charged pion loops, and decuplet mixing) 
from 
short-range (promoted counterterm)
contributions. 
While this would require a dedicated effort, 
longstanding puzzles may be illuminated with future lattice QCD results.

%%%%%%%%%%%%%%%%%%%%%%%%%%%%%%%%%%%%%%%

%%%%%%%%%%%%%%%%%%%%%%%%%%%%%%%%%%%%%%%
\begin{acknowledgments}
%%%%%%%%%%%%%%%%%%%%%%%%%%%%%%%%%%%%%%%
This work was supported in part by the U.S.~National Science Foundation, under Grant No.~PHY$15$-$15738$. 
We would like to thank Johannes Kirscher and members of the $\texttt{NPLQCD}$ collaboration for useful discussions.
%%%%%%%%%%%%%%%%%%%%%%%%%%%%%%%%%%%%%%%
\end{acknowledgments}
%%%%%%%%%%%%%%%%%%%%%%%%%%%%%%%%%%%%%%%

%%%%%%%%%%%%%%%%%%%%%%%%%%%%%%%%%%%%%%%
\appendix
%%%%%%%%%%%%%%%%%%%%%%%%%%%%%%%%%%%%%%%

%%%%%%%%%%%%%%%%%%%%%%%%%%%%%
\section{Magnetic Moments and Electric Polarizabilities}%
\label{s:magpol}%							     %
%%%%%%%%%%%%%%%%%%%%%%%%%%%%%

For completeness, 
we utilize the results of the main text to determine the magnetic moments and electric polarizabilities of the octet baryons to 
$\c O(\epsilon^3)$.
For magnetic moments, 
this represents the next-to-leading order result, 
as tree-level contributions from the operators in  
Eq.~\eqref{eq:colgla}  
scale as 
$\c O(\epsilon^2)$; 
while, 
for electric polarizabilities,
this order constitutes the leading-order result. 
For the latter, 
we determine a value for the electric polarizability of the 
$\S^0$--$\L$
transition, 
a quantity that appears to be overlooked in the literature;
however, 
it is an order of magnitude smaller than the diagonal matrix elements.

To determine magnetic moments, 
notice that the computation of the spin-dependent baryon energies, 
Eq.~\eqref{eq:main}, 
has been renormalized by a subtraction of the zero magnetic field results. 
This regularization-independent subtraction removes the ultraviolet divergences of loop diagrams.
Carrying out the computation of the loop diagrams using dimensional regularization, 
by contrast,  
allows one to renormalize the chiral-limit magnetic moments, 
and thereby determine the corrections away from the chiral limit. 
This is the way in which 
one recovers the known results for chiral corrections to the baryon magnetic moments%
~\cite{Jenkins:1992pi,Durand:1997ya,Meissner:1997hn}, 
namely
\beq \label{eq:magmom}
\mu_B
= 
\a_D \mu_D + Q \mu_F
-
4 M_N
\sum_{\c B}
\c A_\c B \, S_{1\c B} \frac{Q_\phi m_\phi}{(4 \pi f_\phi)^2}
\c F_1
\left( \frac{\D_\c B}{m_\phi} \right)
,\eeq
where the loop function depends on the baryon mass splittings, 
and is given by
\beq
\c F_1(y)
=
\sqrt{y^2 - 1}
\,
\log\left(\frac{y-\sqrt{y^{2} -1+i\e}}{y+\sqrt{y^{2}-1+i\e}}\right)
+
y
\log (4 y^2)
,\eeq
and has been renormalized to vanish in the chiral limit for 
$\D_\c B > 0$,
that is 
$\c F_1(\infty) = 0$. 
Notice further the value of the loop function for vanishing mass splitting,
$\c F_1(0) = \pi$. 
The tree-level coefficients,
$Q$
and
$\a_D$, 
appear in Table~\ref{tab:one}, 
while the loop coefficients, 
$\c A_\c B$, 
and splittings, 
$\D_\c B$,  
are given in 
Table~\ref{tab:two}. 
The factors 
$\c S_{1\c B}$
from the spin algebra are those appearing in 
Eq.~\eqref{eq:main}. 
From a least-squares analysis, 
we fit 
$\mu_D$
and
$\mu_F$
using the experimentally measured magnetic moments, 
and 
$\S^0$--$\L$
transition moment. 
The fit is performed at tree-level, 
and the leading one-loop order. 
Results are provided in Table~\ref{tab:three}, 
and show reasonable agreement with experiment, 
with the exception of the 
$\L$
baryon which differs considerably when the one-loop corrections are taken into account. 
We have also performed fits treating the tree-level computation in baryon magneton units, 
\emph{i.e.}~by considering the Coleman-Glashow operators with a factor
$1/M_B$ 
that depends on the octet baryon state. 
These fits largely show a systematic improvement between tree-level and leading-loop order, 
however, 
the
$\L$
remains problematic. 
For this reason, 
we do not tabulate baryon magneton fit results.

%%%%%%%%%%%%%%%%%%%%%%%%%%%%%%%%%%%%%%%%%%%%%
\begin{table}
\begin{center}
\def\arraystretch{1.25}
\caption{Octet baryon magnetic moments and electric polarizabilities determined using 
$\chi$PT, 
along with experimental values%
~\cite{Olive:2016xmw}.%
\footnote{
The two starred values are derived from experiment with additional assumptions. 
In the case of 
$\mu_{\Sigma^0}$, 
we use the isospin symmetry prediction, 
$\mu_{\S^0} = \frac{1}{2} ( \mu_{\S^+} + \mu_{\S^-})$, 
and assign an additional one-percent uncertainty in quadrature. 
This moment is not included in the fits. 
As only 
$| \mu_{\Sigma^0 \L}|$ 
has been measured, 
moreover,
we take the sign consistent with 
$SU(3)_V$ 
symmetry given that expectations for 
$SU(3)_V$
breaking are 
$\sim 30 \%$
rather than
$\sim 200 \%$.   
The transition moment is included in the fits. 
}
All values for magnetic moments are given in nuclear magnetons, 
while electric polarizabilities are given in units of 
$10^{-4} \, \texttt{fm}^3$.
While 
$\chi$PT 
results have been quoted to two decimal precision, 
this does not necessarily reflect the accuracy of the computed values. 
} 
\label{tab:three}
\medskip
\begin{tabular}{ |L||C|C|C||C|C||  }
\hline
\hline
B
& 
\mu_B: \, \mathcal{O}(\epsilon^2) 
& 
\mu_B: \, \mathcal{O}(\epsilon^3) 
&
\mu_B^{\text{expt.}}
&
\alpha_E: \, \mathcal{O}(\epsilon^3)
&
\alpha_E^\text{expt.}
\tabularnewline
\hline 
p 
& 
\phantom{-}
2.59
& 
\phantom{-} 
3.12
&  
\phantom{-}
2.793(0)
&
11.53
&
11.2(4)
\phantom{1}
\tabularnewline
n  
&
-1.65
&
-2.17
&  
-1.913(0)
&
11.09
&
11.8(11)
\tabularnewline
\Lambda 
&
-0.82
& 
-0.33
&
-0.613(4)
&
\phantom{1}
6.41
&
-
\tabularnewline
\Sigma^{+}
& 
\phantom{-}
2.59 
& 
\phantom{-}
2.38 
& 
\phantom{-i}
2.458(10)
&
11.26
&
-
\tabularnewline
\Sigma^{0}
& 
\phantom{-}
0.82 
& 
\phantom{-}
0.61 
& 
* \, \, \, 0.649(15)
&
\phantom{1}
7.69
&
-
\tabularnewline
\Sigma^{-} 
&
-0.94
&
-1.15
&
-1.160(25)
&
\phantom{1}
9.06
&
-
\tabularnewline
\Xi^{0} 
&
-1.65 
& 
-0.84
& 
-1.250(14)
&
\phantom{1}
4.68
&
-
\tabularnewline
\Xi^{-} 
& 
-0.94 
& 
-0.41
& 
-0.650(3)
\phantom{1}
&
\phantom{1}
2.72
&
-
\tabularnewline
\Sigma^{0}\rightarrow\Lambda 
& 
\phantom{-}
1.43 
& 
\phantom{-}
1.65 
&
* \, \, \,
1.61(8)
\phantom{11}
&
-0.45 
\phantom{i}
&
-
\tabularnewline
\hline
\hline
\end{tabular}
\end{center}
\end{table}
%%%%%%%%%%%%%%%%%%%%%%%%%%%%%%%%%%%%%%%%%%%%%

The baryon electric polarizabilities can be determined in $\chi$PT.  
One way to obtain the electric polarizabilities of the octet baryons is to determine the energy levels in the presence of a weak uniform electric field. 
Within the heavy baryon approach, 
the acceleration of charged baryons does not become relevant in loop diagrams until 
$\c O(\epsilon^4)$. 
Thus, 
using the procedure outlined in 
Ref.~\cite{Tiburzi:2008ma}, 
we obtain the standard expression for meson loop contributions to the electric polarizability,
given by
\beq \label{eq:Epols}
\alpha_{E}
=
\frac{\alpha}{3}
\sum_\c B
\frac{\c A_\c B \, \c S_{2 \c B}}{m_\phi (4 \pi f_\phi)^2}
\c F_2 \left( \frac{\D_\c B}{m_\phi} \right),
\eeq
where
$\alpha$
is the fine-structure constant, 
the 
$\c A_\c B$
coefficients are tabulated in 
Table~\ref{tab:two}, 
the spin factors, 
$\c S_{2 \c B}$, 
are those in 
Eq.~\eqref{eq:main}, 
and the loop function is given by
\beq
\c F_2(y)
=
\frac{9\, y}{y^2 - 1}
-
\frac{y^2 -10}{2 (y^{2}-1)^{\frac{3}{2}}}
\log\left(
\frac{y-\sqrt{y^{2}-1+i\e}}
{y+\sqrt{y^{2}-1+i\e}}
\right)
,
\eeq
which has the particular value
$\c F_2(0) = 5 \pi$.
The pion loop contributions to these results agree with those in
Ref.~\cite{Bernard:1991rq,Butler:1992ci,Hemmert:1996rw} 
for nucleon electric polarizabilities, 
and 
the octet loop contributions to hyperon electric polarizabilities agree with those determined in 
Ref.~\cite{Bernard:1992xi}.
The electric polarizabilities obtained from 
Eq.~\eqref{eq:Epols} 
are collected in Table \ref{tab:three}. 
The nucleon electric polarizabilities determined in three-flavor $\chi$PT agree well with experiment, 
and this is to be expected given the dominance of the pion loop contributions. 
The small difference between proton and neutron polarizabilities is attributable to differing kaon loop contributions, 
and is comparable to the experimental uncertainties. 
The previously overlooked transition polarizability between the 
$\S^0$
and
$\L$
baryons is predicted to be negative and an order of magnitude smaller than the diagonal polarizabilities in this system. 
This smallness occurs because only the kaon loops contribute.

%%%%%%%%%%%%%%%%%%%%
\section{Coupled $I_3 = 0$ Baryons}
\label{s:B}
%%%%%%%%%%%%%%%%%%%%

%%%%%%%%%%%%%%%%%%%%%%%%%%%%%%%%%%%%%%%
\begin{figure}
\centering
\resizebox{\linewidth}{!}{
        \includegraphics[width=6cm]{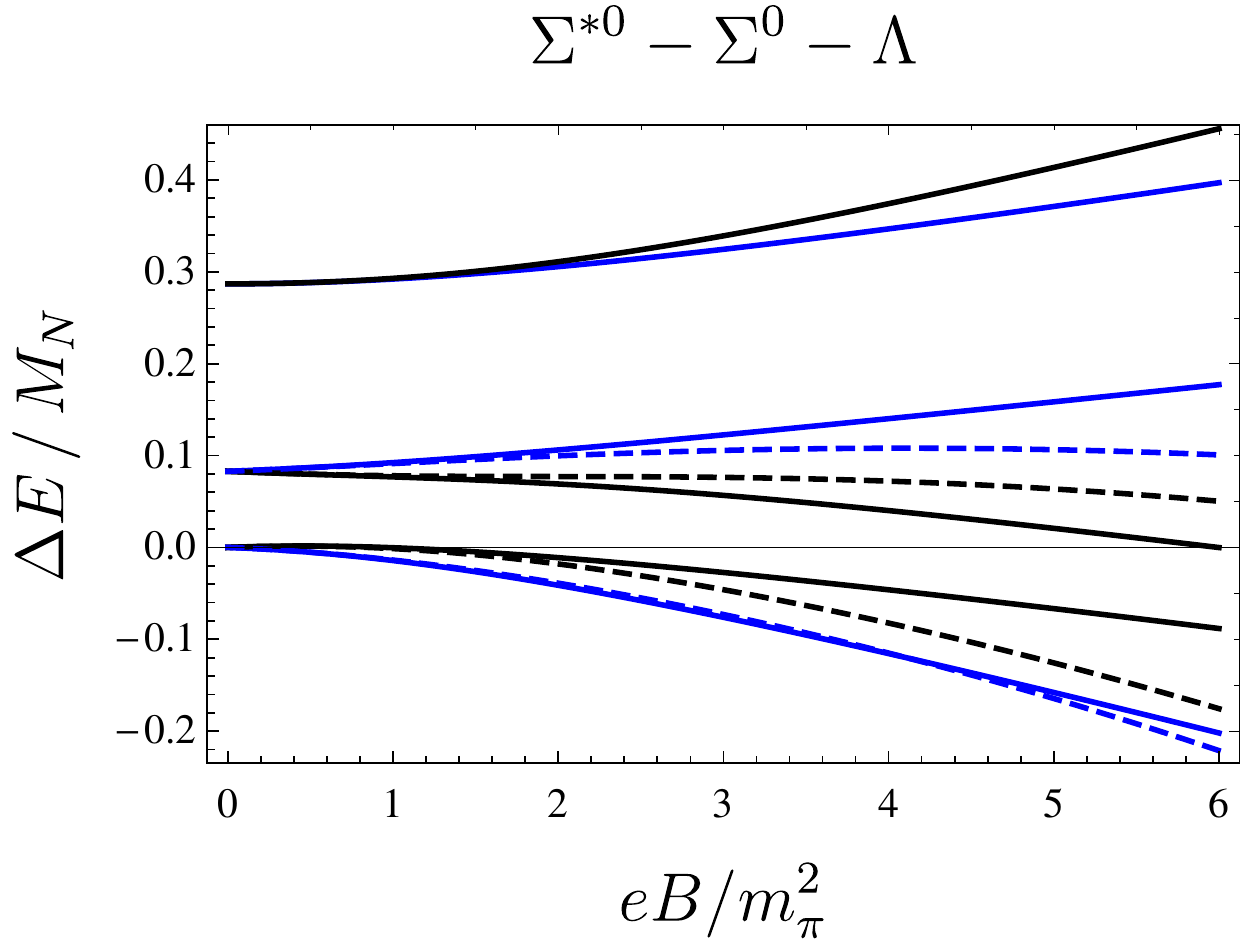}
        $\quad$
        \includegraphics[width=6cm]{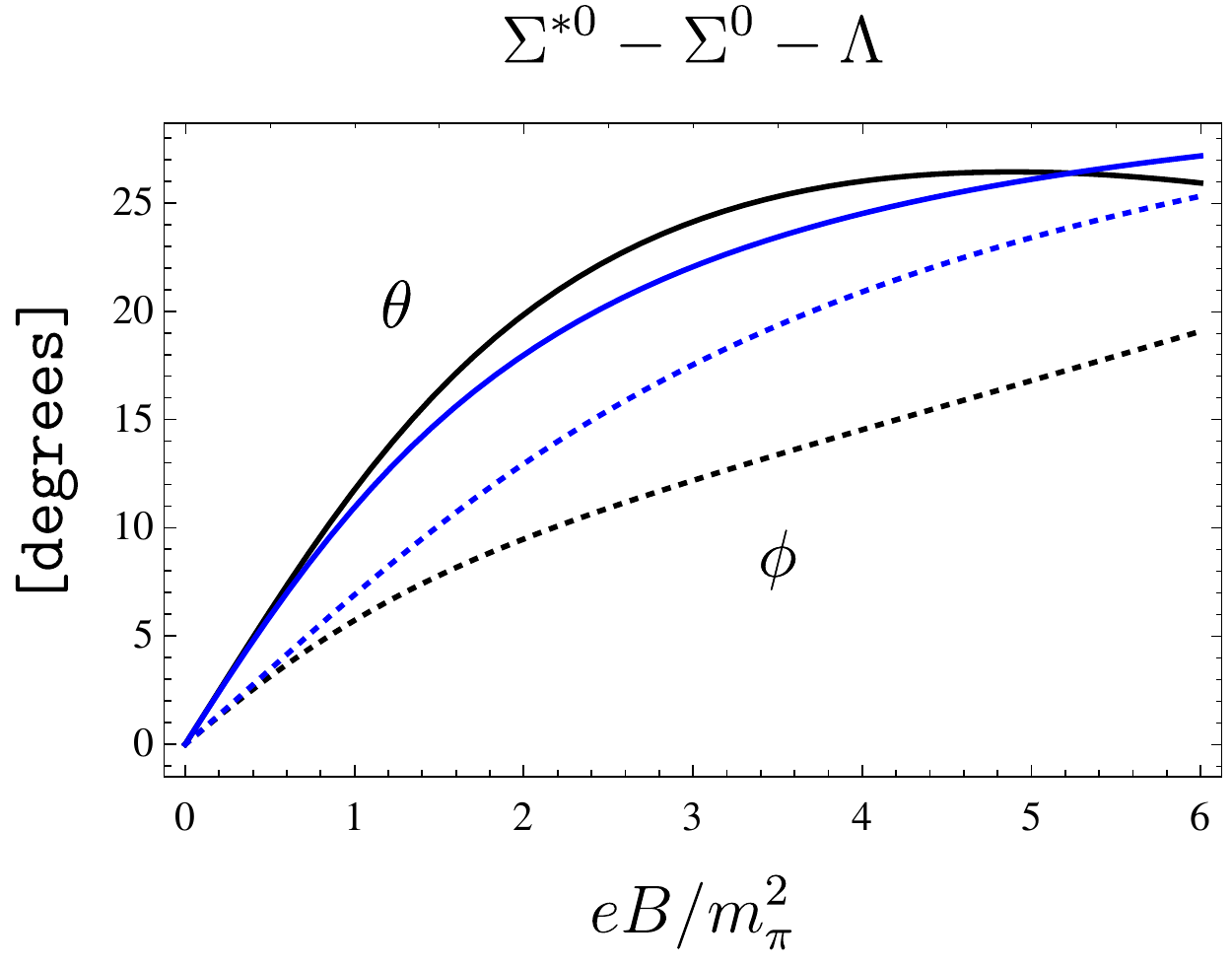}
}
%%%%%%%%%%%%%%%%%%%%%%%%%%%%%%%%%%%%%%%
\caption{
Assessment of the magnetic field dependence of eigenstate energies in the coupled 
$\S^{*0}$--$\S^0$--$\L$ 
system. 
On the left, 
solid curves correspond to the energy shifts, 
$\D E = E - M_\L$,  
of the eigenstates of 
Eq.~\eqref{eq:3mix}, 
which treats the magnetic moments of the 
$I_3 = 0$ 
baryons to all orders, 
with black curves for spin-up states and blue curves for spin-down states. 
The dashed curves correspond to the energies obtained by treating mixing with the 
$\S^{*0}$
baryon perturbatively, 
according to
Eq.~\eqref{eq:2mix}.  
On the right, 
the mixing angles 
$\theta$
and
$\phi$
of 
Eq.~\eqref{eq:angles}
are plotted as a function of the magnetic field, 
with black for spin-up states and blue for spin-down states. 
}
        \label{f:3state}
\end{figure}
%%%%%%%%%%%%%%%%%%%%%%%%%%%%%%%%%%%%%%%

In large magnetic fields, 
the size of the baryon transition moment, 
$\mu_U$, 
may lead to non-perturbative mixing between decuplet and octet baryons. 
Here, 
we consider the effects of mixing in the coupled system of three 
$I_3 = 0$ 
baryons. 
Accounting for the magnetic moment interactions, 
this system is described by the Hamiltonian
\beq
H
=
\begin{pmatrix}
M_{\S^{*}} 
& 
0 
& 
0 
\\
0 
& 
M_{\S} 
& 
0 
\\
0 
& 
0 
& 
M_\L
\end{pmatrix}
-
\frac{e B}{2 M_N}
\begin{pmatrix}
2 m \, \mu_{\S^{* 0}} 
& 
 \mu_{\S^{*0} \S^0}
& 
\mu_{\S^{*0} \L}
\\
\mu_{\S^{*0} \S^0}
& 
2 m \, \mu_{\S^0} 
& 
2 m \mu_{\S^0 \L}
\\
\mu_{\S^{*0} \L}
& 
2 m \mu_{\S^0 \L}
& 
2 m \, \mu_\L
\end{pmatrix}
\label{eq:3mix}
,\eeq
which has been written in the basis 
$\begin{pmatrix} \S^{* 0} \\ \S^0 \\ \L \end{pmatrix}$, 
with 
$m = \pm \frac{1}{2}$
denoting the spin projection along the magnetic field. 
To simplify the analysis slightly, 
we take 
$\mu_{\S^{* 0}} = 0$, 
which is the $U$-spin prediction and consistent with nearly all calculations, 
see
Ref.~\cite{Geng:2009ys}. 
The magnetic moment of the 
$\L$
baryon is taken as the experimental one, 
and the starred experimental values of 
Table~\ref{tab:three} are used for
$\mu_{\S^0 \L}$
and
$\mu_{\S^0}$.
The baryon transition moments are given in terms of the $U$-spin symmetric coefficient 
$\mu_U$
as:
$\mu_{\S^{*0} \S^0} = - \frac{1}{2 \sqrt{3}} \mu_U$
and
$\mu_{\S^{*0} \L} = \frac{1}{2} \mu_U$. 
To assess the size of mixing in this three-state system, 
we adopt the numerical value for 
$\mu_U$
determined in 
Sec.~\ref{s:mixy}.  
While the energy eigenvalues of the Hamiltonian in 
Eq.~\eqref{eq:3mix} 
can be determined as the roots of a cubic polynomial, 
the analytic expressions offer little insight. 
The magnetic field dependence of the eigenstate energies of 
Eq.~\eqref{eq:3mix} is shown in 
Fig.~\ref{f:3state}. 
For comparison, 
we determine the energies with 
$\S^{*0}$
mixing treated perturbatively. 
This approximation is described by the Hamiltonian
\beq
H
=
\begin{pmatrix}
M_\S 
- 
\frac{\mu_{\S^{*0} \S^0}^2}{M_{\S^*} - M_\S}
\left( \frac{e B}{2 M_N} \right)^2
& 
0
\\
0 
&
M_\L 
-
\frac{\mu_{\S^{*0} \L^0}^2}{M_{\S^*} - M_\L}
\left( \frac{e B}{2 M_N} \right)^2
\end{pmatrix}
-
m\frac{e B}{ M_N}
\begin{pmatrix}
 \mu_{\S^0} 
& 
 \mu_{\S^0 \L}
\\ 
 \mu_{\S^0 \L}
& 
 \mu_\L
\end{pmatrix}
\label{eq:2mix}
,\eeq
which accounts for mixing among 
$\S^0$
and 
$\L$, 
but only treats the 
$\S^{*0}$
through the pole diagram, 
see Fig.~\ref{fig:two}. 
For the value of 
$\mu_U$
employed, 
mixing with the 
$\S^{*0}$
is seen to be appreciable for 
$e B / m_\pi^2 \gtrsim 2$. 
A way to write spin-up eigenstate of lowest energy,
$| \lambda_- \rangle$,
is in terms of mixing angles 
$\theta$
and 
$\phi$
defined through the relation
\beq
| \lambda_- \rangle
=
\cos \phi 
\,
\cos \theta 
\, 
| \L \rangle
+ 
\cos \phi 
\,
\sin \theta 
\, 
| \S^0 \rangle
+
\sin \phi \, | \S^{*0} \rangle
\label{eq:angles}
.\eeq
For the spin-down eigenstate of lowest energy, 
we use the opposite
sign phase convention for $\theta$, which accounts for the sign flip in the off-diagonal $\Sigma^0$--$\Lambda$ matrix elements.
In this decomposition, 
no mixing with the
$\S^{*0}$
baryon corresponds to 
$\phi = 0$, 
for which
$\theta$
is the mixing angle of 
Eq.~\eqref{eq:lamsigmix}. 
In 
Fig.~\ref{f:3state},
we show the magnetic-field dependence of the mixing angles determined from the eigenvector 
$| \l_- \rangle$
of 
Eq.~\eqref{eq:3mix}.  
The figure shows that mixing with the 
$\S^{* 0}$
baryon is expected to be almost as important as 
$\S^0$--$\L$ 
mixing.

In Sec.~\ref{s:energies} of the main text, 
we move beyond the above assessment of 
$\S^{*0}$--$\S^0$--$\L$
mixing to consider additionally the magnetic polarizabilities and charged-meson loop effects. 
The results shown in 
Fig.~\ref{fig:dif9} 
are obtained from the lowest two eigenvalues of the Hamiltonian
\beq
H
=
\begin{pmatrix}
M_{\S^*}
& 
0 
& 
0 
\\
0 
&
\ol E_{\S^0}
&
\ol E_{\S^0 \L}
\\
0
&
\ol E_{\S^0 \L}
&
\ol E_\L
\end{pmatrix}
- 
\frac{e B}{2 M_N}
\begin{pmatrix}
2 m \, \mu_{\S^{*0}} 
&
\mu_{\S^{*0} \S^0}
&
\mu_{\S^{*0} \L}
\\
\mu_{\S^{*0} \S^0}
&
2 m \, \c M_{\S^0}
&
2 m \c M_{\S^0 \L}
\\
\mu_{\S^{*0} \L}
&
2 m \c M_{\S^0 \L}
&
2 m \, \c M_{\L}
\end{pmatrix}
\label{eq:3state}
,\eeq
where the spin-independent entries,
$\ol E_B$, 
are given in 
Eq.~\eqref{eq:Ebar}, 
and the spin-dependent entries, 
$\c M_B$, 
are given in 
Eq.~\eqref{eq:MBs}.
In contrast with 
Eq.~\eqref{eq:3mix}, 
the transition moments in 
Eq.~\eqref{eq:3state} 
are taken to be the values rescaled by 
$\sqrt{\gamma}$, 
with 
$\gamma$ 
given in Eq. (49).  
These eigenvalues are compared with those obtained from retaining all 
$\c O (B^2)$ 
contributions to the two-state problem, 
i.e.~eigenvalues of 
\beq
H
=
\begin{pmatrix}
M_\S
+ 
\D E_{\S^0}
& 
\D E_{\S^0 \L}
\\
\D E_{\S^0 \L}
&
M_\L 
+ 
\D E_\L
\end{pmatrix}
\label{eq:2state}
,\eeq
where the 
$\D E$ 
matrix elements are given by 
Eq.~\eqref{eq:B2}.

%%%%%%%%%%
\bibliography{bibly}%
%%%%%%%%%%%

%%%%%%%%%%%%%%%%%%%%
\end{document}